\documentclass{wp}

\usepackage[comma]{natbib}
\usepackage[hyphens]{url}
\usepackage{hyperref}
\usepackage{breakurl}

\hypersetup{
    breaklinks=true,
    colorlinks=true,
}

\jname{Xxxx. Xxx. Xxx. Xxx.}
\jvol{AA}
\jyear{YYYY}
\doi{10.1146/(n.a.)}

\begin{document}

\markboth{Abowd \& Hawes}{2020 Census Confidentiality}

\title{Confidentiality Protection in the 2020 US Census of Population and Housing}

\author{John M. Abowd and Michael B. Hawes 
\affil{US Census Bureau, Washington, DC, USA; email: john.maron.abowd@census.gov, michael.b.hawes@census.gov}
}

\begin{abstract}
In an era where external data and computational capabilities far exceed statistical agencies' own resources and capabilities, they face the renewed challenge of protecting the confidentiality of underlying microdata when publishing statistics in very granular form and ensuring that these granular data are used for statistical purposes only. Conventional statistical disclosure limitation methods are too fragile to address this new challenge. This article discusses the deployment of a differential privacy framework for the 2020 US Census that was customized to protect confidentiality, particularly the most detailed geographic and demographic categories, and deliver controlled accuracy across the full geographic hierarchy. 
\end{abstract}

\maketitle

\tableofcontents

\section{INTRODUCTION}
The US Census of Population and Housing has Constitutional origins, statutory uses, and millions of diverse applications that are difficult to comprehensively catalogue. Fifty counts---the resident population plus the federally affiliated overseas population of each state---are used to apportion the House of Representatives (2 U.S.C. § 2a). The Redistricting Data Summary File (P.L. 94-171, 13 U.S.C. § 141) is used to assist states, the District of Columbia, and the Commonwealth of Puerto Rico in redrawing legislative districts at all levels of government and to allow scrutiny of those districts for prohibited racial and ethnic discrimination under the 1965 Voting Rights Act (VRA). The census data are then used for the Population Estimates Program (including uses by the Federal-State Cooperative for Population Estimates), whose output is subsequently used for the population controls in many of the agency's household surveys, including the American Community Survey. The allocation of federal, state, and municipal resources resulting from these and other downstream uses is most comprehensively summarized by Reamer (2020). 

To accommodate these very diverse applications, the geographic detail and tabular complexity of decennial census publications have grown enormously over the decades. This expansion of published output culminated in the tabular and microdata summaries for the 2010 Census, which released approximately 150 billion statistics (counting the zeros in tables that were entirely empty) based on about 15 billion bytes of data or about 1.5 billion distinct attributes (Garfinkel et al. 2019).\footnote{The total number of statistics released from the 2010 Census, as reported by Garfinkel et al. (2019) was later revised to the number reported here based on an exhaustive accounting of Summary Files 1 and 2, the  American Indian/Alaska Native Summary File and the Public-Use Microdata Sample (PUMS) conducted by the statisticians who implemented the 2020 Census Disclosure Avoidance System.} Many of these tabulations were produced at the census-block level---a geographic unit with an average population of 29 persons in 2010---primarily in support of drawing new voting districts after the data are released and producing summaries for many distinct legal and statistical areas.

At the same time, the problem of protecting confidentiality, a consistent statutory requirement since 1954, became increasingly difficult. The Census Bureau's Data Stewardship Executive Policy Committee (DSEP) acknowledged this conundrum when it approved the release of the 2010 Census data products, saying that ``the problem of block population size and disclosure avoidance is real...it deserves attention in the 2020 [Census] planning'' (DSEP 2010, p. 2). The National Institutes of Health (NIH) faced the same dilemma when biostatisticians demonstrated that genome-wide association studies (GWAS) were vulnerable to reidentification attacks (Homer et al. 2008). Modern privacy-preserving publication systems acknowledge the tension between data accuracy, the volume of statistics released, and confidentiality protection. They deliberately work from leading use cases rather than attempting to publish statistics that meet nearly all known uses. This article is about these confidentiality protections, known in the United States as statistical disclosure limitation (SDL), which is known as disclosure avoidance at the Census Bureau for purely historical reasons. 

This article is organized as follows. In Section \ref{sec:history}, we provide a brief history of the evolution of the American population census. Section \ref{sec:changed} describes how the information technology environment has  changed since the development of the original disclosure avoidance methods in the 1970s. The uses of statistical data also changed, as we describe in Section \ref{sec:tabulations}. To accommodate the changing environment and uses, computer scientists and statisticians developed methods for quantifying global disclosure risk, which we discuss in Section \ref{sec:accounting}. Then, we review the alternatives available for the 2020 Census in Section \ref{sec:alternatives}. In Section \ref{sec:DAS}, we provide the fundamentals of the disclosure avoidance system used for the 2020 Census.  Our use-case discussion centers on redistricting, which we elaborate in Section \ref{sec:uses}. The challenge of appropriate scientific inference when using differential privacy (DP) frameworks is discussed in Section \ref{sec:inference}. Finally, we draw some lessons for the future in Section \ref{sec:future}.

\section{A BRIEF HISTORY OF CENSUS CONFIDENTIALITY}
\label{sec:history}

To secure voluntary participation, US Marshals stopped posting household enumerations in the town square in the 1850 Census. Congress passed the first confidentiality protections for individual census responses in the Census Act of 1879. These efforts were expanded and codified in 13 U.S.C. § 8(b) and 9, which prohibit the Census Bureau from releasing ``any publication whereby the data furnished by any particular establishment or individual under this title can be identified[,]'' and allows the release of aggregate statistics so long as those data ``do not disclose the information reported by, or on behalf of, any particular respondent[.]'' The Confidential Information Protection and Statistical Efficiency Act of 2018 (CIPSEA) requires all statistical agencies to ``protect the trust of information providers by ensuring the confidentiality and exclusive statistical use of their responses'' (44 U.S.C. § 3563).

The broader scientific community concurs with the importance of rigorous confidentiality protection by statistical agencies. For example, the National Academies of Sciences, Engineering, and Medicine (NASEM) publish the definitive guidebook for federal statistical agencies, which states that ``[b]ecause virtually every person, household, business, state or local government, and organization is the subject of some federal statistics, public trust is essential for the continued effectiveness of federal statistical agencies. Individuals and entities providing data directly or indirectly to federal statistical agencies must trust that the agencies will appropriately handle and protect their information'' (NASEM 2021, p. 3).  The report also notes that respondents expect statistical agencies not to ``release or publish their information in identifiable form'' (NASEM 2021, p. 36).  The NASEM also broadly exhorts statistical agencies to ``continually seek to improve and innovate their processes, methods, and statistical products to better measure an ever-changing world'' (NASEM 2021, p. 4).

Public trust that the Census Bureau will keep data confidential is essential for respondent cooperation. There are extensive outreach efforts to reassure respondents and other data providers of the agency’s commitment to  protecting confidential data. The criminal fines and imprisonment penalties that the agency's employees would face by unlawfully disclosing respondent information are frequently cited in these outreach efforts (USCB 2019). The connection between response rates and respondent perception of confidentiality has been continuously studied, even as the 2020 Census was starting fieldwork (Childs et al. 2020).

The modern decennial census format (self-response) and publications (massive tabular summaries) date from the 1960 Census. Routine publication of block-level summaries began with modifications to the Census Act passed in 1975 (P.L. 94-171, 13 U.S.C. § 141). Low-level geographic detail was already an acknowledged confidentiality concern in 1930. Prior to the 1990 Census, the primary mechanism that the agency employed to protect the confidentiality of census responses was to suppress any table that did not meet certain household, population, or demographic characteristic thresholds. The 1970 Census publications, for example, suppressed tables reflecting fewer than five households and only included tables of demographic characteristics cross-tabulated by race if there were at least five individuals in each reported race category (McKenna 2018, p. 3). These suppression routines helped to protect confidentiality by reducing the published detail about individuals who were relatively unique within their communities. 

For the 1990 Census, however, the Census Bureau replaced suppression for two reasons: first, data users were dissatisfied with missing details caused by suppression and second, the agency realized that the suppression it had been using was insufficient to fully protect against reidentification (McKenna 2018, p. 6). To replace suppression, the agency introduced noise infusion to safeguard confidentiality.  \textit{Noise infusion} introduces controlled amounts of deliberate error or noise into the published tables, where the method of control depends upon the technique used. Noise infusion tries to preserve the overall statistical validity of published data while introducing enough uncertainty that attackers would not have any reasonable degree of certainty that they had correctly isolated data for any particular respondent.  The noise infusion used in 1990 was a simple form of data swapping between paired households in a geographic area with similar attributes. Additionally, for small block groups, the Census Bureau replaced the collected characteristics of households with imputed characteristics (McKenna 2018, p. 6)---a method called \textit{blank and impute}, which was only used in the 1990 Census (McKenna 2108, p. 11). Importantly, the 1990 redistricting data were unaltered by the swapping algorithm because all variables in those tables were part of the swap key (McKenna 2018, p. 6).

For the 2000 and 2010 Censuses, the Census Bureau infused noise using more targeted data swapping criteria. The agency first identified households most vulnerable to reidentification---especially households on smaller-population blocks and those whose residents had differing demographic characteristics from the remainder of their block. While every nonimputed household record\footnote{When a respondent household provides only a count of the number of persons living at that address or when the housing-unit population count is itself imputed, the Census Bureau imputes all characteristics: sex, age, race, ethnicity, and relationship to others in the household. Such persons are called whole-person census imputations in technical documentation. When a household consists entirely of whole-person census imputation records, it is called an imputed household. A nonimputed household contains at least one person whose characteristics were collected on the census form for that household.} in the Census Edited File (CEF) had a chance of being selected for data swapping, records for more vulnerable households were selected with greater probability (McKenna 2018, p. 8). Then, the records for all members of those selected households were exchanged with the records of households in a nearby geographic area that matched on key characteristics called the swap key.\footnote{The definition of ``nearby'' is left deliberately vague in technical documentation because it is considered a confidential feature of the data swapping algorithm.} For the 2000 and 2010 Censuses, those key matching characteristics were (\textit{a}) the whole number of persons in the household and (\textit{b}) the whole number of persons aged 18 or older in the household. These swapping criteria resulted in the total population and total voting age population for each block being held invariant—that is, while noise was added to all remaining characteristics, no noise was added to the block-level total population or block-level voting age population counts. For the first time, in the 2000 Census publications, the data in the redistricting files at the block level were noisy versions of the same tabulations from the CEF. 

Any discussion of swapping must take note of the Census Bureau's confidentiality protection semantic claim: ``Because of data swapping, users should not assume that tables with cells having a value of one or two reveal information about specific individuals'' (USCB 2011, section 7, p. 6). The claim applies to tabular data. By contrast, in the technical documentation for the 2010 Census PUMS, minimum population sizes are cited as the primary confidentiality protection: ``The main disclosure avoidance method used is to limit the geographic detail shown in the files. A geographic area must have a minimum population of 100,000 to be fully identified. A minimum threshold of 10,000 for the national population (excluding Puerto Rico) was set for identification of groups within categorical variables in the state-level PUMS files'' (USCB 2015, section 2, p. 1). The different standards are explicitly catalogued by McKenna (2018, 2019). The mutual inconsistency of these claims is evident to any modern analyst of privacy protection and is elaborated in Section \ref{sec:changed}. The claim asserts that the protection afforded by swapping extends to tables based on geographic units with populations as small as 1 and averaging 29 persons for all blocks in 2010, but in microdata format, swapping is inadequate unless the minimum population of the geographic area is at least 100,000 (the minimum population of a Public-Use Microdata Area) and the minimum national population for the cells of categorical variables (e.g., race and ethnicity) is 10,000. The fundamental assumption was that tabular summaries were qualitatively different from microdata in their confidentiality protection requirements. This assumption is pervasive in traditional SDL methods (e.g., Duncan et al. 2011, chapter 4 on tabular data and chapter 5 on microdata), and it is provably false (Dinur \& Nissim 2003, Dwork \& Yekhanin 2008).

\section{HOW THE WORLD AROUND THE CENSUS BUREAU CHANGED}
\label{sec:changed}


In the twenty-first century, it has become clear that long-held assumptions about the confidentiality risks associated with highly granular data publication are demonstrably false. First, large-scale equation-solving software means that database reconstruction of the confidential data from tabular publications is feasible and practical. Second, massive record-level external databases are common in nonstatistical government agencies and private companies. These two developments seal the fate of the SDL methods in widespread use for population censuses---swapping and aggregation (Abowd 2017, 2021a; JASON 2020, pp. 34--40).

Aggregation, especially over many data points, has long been the most basic confidentiality protection method. Instead of posting the list of all census responses in the town square, censuses in  second-half of the nineteenth century published the count of persons who live in the village. In other disciplines, aggregation was also considered sufficient confidentiality protection. For example, as human genomes became available, the NIH recognized the inherent uniqueness of individual genetic data and put consent and confidentiality safeguards in place for the genomes themselves, but it permitted the publication of GWAS with neither prior consent of the case study participants nor explicit confidentiality protections beyond aggregation. Homer et al. (2008) showed that with reliable auxiliary data on the control population for a GWAS, a single genome could be reliably classified as either case study or control. The precision of the classifier depended upon the number of single nucleotide polymorphisms (SNP) tabulated in the GWAS and the volume of control data. They concluded that ``these findings also suggest that composite statistics across cohorts, such as allele frequency or genotype counts, do not mask identity within genome-wide association studies'' (Homer et al. 2008, p. 1).

Dwork et al. (2015, p. 3) showed that to build a classifier with nearly perfect precision ``it suffices for the attacker to have a single reference sample from the population.'' That is, they showed that if an attacker has a single genome from the reference population (control) and wants to classify another genome as either case study or control, then, under the assumptions of independence of the alleles, independence of the control samples, and number of alleles at least $O(n^2 \log(1/\delta))$, the classifier's precision---probability of correct classification---is at least $1-\delta$. They also proved that the precision of the reidentification classifier could be directly controlled by publishing the GWAS using DP on the SNP frequencies. Yu et al. (2014) provided a full DP implementation for the $\chi^2$ statistics associated with the SNP frequencies in the GWAS.

The NIH did not wait for the detailed mathematics or the full solution. On August 25, 2008---just weeks after the Homer et al. (2008) article appeared---it announced that ``new statistical techniques for analyzing dense genomic information make it possible to infer the group assignment (i.e., case or control) of an individual DNA sample'' (NIH 2008, pp. 1-2). From that date, an approved study needed informed consent to use a genome in a public GWAS, and lacking such consent, the GWAS became a restricted-access data item.

The NIH's decision to restrict access to GWAS is an early example of modifying confidentiality protections because of feasible reidentification attacks on aggregate data. 
The release of any statistic calculated from a confidential data source will reveal a potentially trivial, but nonzero, amount of confidential information. 
If an attacker has access to enough aggregated data with sufficient detail and precision, then the attacker may be able to leverage information from each statistic in the aggregated data to reconstruct the individual-level records that were used to generate the published tables (Dinur \& Nissim 2003, Dwork \& Yekhanin 2008). This process is known as a \textit{reconstruction attack}, and it adds a new degree of disclosure vulnerability against which statistical agencies must defend. While the statistical and computer science communities have been aware of this vulnerability since 2003, the computing power and sophisticated numerical optimization software necessary to perform these reconstructions appeared later in the 2000s.

Statistical releases need not all be of the same type or contain the same data fields to enable reidentification by reconstruction. For example, a 2015 interagency report published by the National Institute of Standards and Technology provided many examples of using disparate data sets to reconstruct hidden underlying data (Garfinkel 2015). These include articles by Narayanan \& Shmatikov (2008), De Montjoye et al. (2013, 2015), and Ma et al. (2013). The same general principles apply to census data.  The difference between census data and the examples above is that census data can be combined in vastly more ways with other information because all tables published from census data share standardized identifiers including location, age, sex, race, ethnicity, and marital status. Even if each of these identifiers is not included in every table, their use and combinations across many different tables create the disclosure risk. 

The risk of disclosure in decennial census data is amply demonstrated by the reconstruction attack the Census Bureau itself simulated (Abowd 2018; J.M. Abowd et al., manuscript under review).\footnote{The first public presentation of the methodology used for the reconstruction portion of the attack was provided by Abowd (2018). Reidentification components were developed over the next several years and are documented by J.M. Abowd et al. (manuscript under review). Herein, only publicly released components of the newer paper are used, and citations are provided to the public versions.} This simulated attack used only a small portion of the publicly released tables from the 2010 Census.\footnote{The full list of tables used in the reconstruction is provided by Hawes (2022).}
We summarize the results using exact/binned age, defined as single year of age for ages 0-21; for ages 22-84, age is binned in 2-5 year increments ([22,24], [25,29], [30,34], [35,39], [40,44], [45,49], [50,54], [55,59], [60,61], [62,64], [65,66], [67,69], [70,74], [75,79], [80,84]), and individuals age 85 or older are top-coded as 85+. The results of this reconstruction were alarming. The agency was able to reconstruct a highly accurate image of the confidential CEF microdata from this small subset of the published 2010 tabulations. As Table 1 
shows, the overall agreement rate [the percentage of reconstructed records that exactly agree with the CEF on location, sex, age (exact/binned), race, and ethnicity] was 91.8\%. Agreement rates were slightly lower (74.0\%) for blocks with 1-9 people, as households in these smallest blocks were the primary target of the 2010 swapping algorithm, but slightly larger blocks with 10-49 people had an agreement rate of 93.0\%. Although the agreement rate for blocks of size 1-9 is lower than for larger-population blocks, the fact that household swapping was used raises a further possibility (not explored) that by considering neighboring blocks and using an external data set containing household age/sex data, swapping might even be undone, so the 74\% does not indicate that swapping was effective in protecting households.

\begin{table}[h]
\label{table:agreement:block:size}
\tabcolsep7.5pt
\caption{Agreement rates (reconstruction to CEF) by block size}
\begin{center}
\begin{tabular}{@{}r|c|c|c|c|c|c|c|c@{}}
\hline
Block size &Total & 1-9 &10-49 &50-99 & 100-249 &250-499 &500-999 &1,000+\\
\hline
Agreement &91.8\% &74.0\% &93.0\% &93.1\% &92.1\% &91.3\% &90.6\% &91.5\%\\
\hline
\end{tabular}
\end{center}
\begin{tabnote}
\begin{flushleft}DRB clearance number CBDRB-FY22-DSEP-004. Data are from J.M. Abowd et al. (manuscript under review), released in Hawes (2022). Abbreviations: CEF, Census Edited File; DRB, Disclosure Review Board.
\end{flushleft}
\end{tabnote}
\end{table}


The team that demonstrated this vulnerability stopped after reconstructing person-level records for block, sex, age, race, and Hispanic ethnicity because the vulnerability had been fully exposed mathematically and demonstrated empirically. The reconstruction experiment confirmed that existing technology can convert the Census Bureau’s traditional tabular summaries into a 100\% coverage microdata file geocoded to the block level with very limited noise. Such a file was not released in 2010. This reconstructed microdata file contains so much detail that it would have been deemed unreleasable if it had been proposed in conjunction with the original 2010 Census data products. When DSEP was presented with the results of this simulated reconstruction attack, the committee immediately realized that the established swapping mechanism resulted in far less disclosure protection than it was intended to provide. The committee decided that disclosure avoidance for the 2020 Census would have to be strengthened (DSEP 2017).


In parallel with the Census Bureau's efforts to improve disclosure avoidance for the 2020 Census, the researchers continued to evaluate the disclosure vulnerability of the prior decades' swapping mechanism. This effort better informed DSEP decision-making about the implementation of disclosure avoidance for the 2020 Census and informs future efforts to modernize disclosure avoidance for other products that have  relied on swapping for confidentiality protection. While successful reconstruction of the underlying microdata from the 2010 Census published tables demonstrated that the agency could no longer treat disclosure avoidance for tabular data products differently than for microdata releases, reconstruction itself does not constitute statistical disclosure. To reidentify the characteristics of particular individuals, an attacker would need to link pseudoidentifiers from the reconstructed microdata to an external file that contains individuals' names or other unique identifiers. The relative success of such a reidentification attack will, of course, depend on the quality and coverage of the external data being used. While the Census Bureau performed an initial simulated reidentification attack using four commercially available data sources from the 2010 time period, higher-quality commercial data with better coverage than the ones the agency used were available in 2010. External data have continued to improve in quality and quantity over the intervening years. In particular, potential attackers have access to many private databases that are not in the public domain. Thus, to establish a worst-case baseline of reidentification risk, the Census Bureau also performed reidentification attacks using the CEF as the external matching file.\footnote{While using the internal CEF as the external matching file constitutes a worst-case baseline in terms of the potential quality of the matching data, this simulated reidentification attack is not a true worst-case disclosure risk analysis. The reconstructed records being reidentified could be enhanced if additional published tables were incorporated into the reconstruction. Furthermore, this simulation is just one possible vector of attack.}   

One of the principal sources of disclosure risk in a reidentification attack is the inherent uniqueness of individuals within their communities. Elamir \& Skinner (2004, p. 2) affirmed this, stating that ``a measure of disclosure risk is the proportion of individuals in the microdata sample which have a unique combination of values of the key variables (assumed categorical) in the population\dots[s]uch individuals, referred to as population unique, may be judged to be particularly `at risk of disclosure'.'' The Federal Committee on Statistical Methodology (FCSM) has also cited this research and explicitly recognized the risk to confidentiality from database reconstruction in their recently released Data Protection Toolkit (\url{https://nces.ed.gov/fcsm/dpt/content/1-3-2}). 

Overall, 44\% of the entire US population has a unique combination of sex and single year of age at the  block level. In sparsely populated blocks, those with populations of between 1 and 9 residents, over 95\% of individuals are population uniques on sex and single year of age at the block level (Abowd 2021b, table 5). The prevalence of population uniques suggests that the underlying disclosure risk of data published to the block level is alarmingly high, and the results of the Census Bureau’s reconstruction-abetted reidentification attack (as reflected in the precision rates of those reidentifications) confirm this vulnerability. The precision of the agency’s simulated attack, shown in Table 2, demonstrates that disclosure risk inherent to the reconstructed data is particularly high for respondents with unique combinations of block, age, and sex. If attackers knew a respondent’s census block, age, and sex and used the reconstructed data to infer their race and ethnicity, they would have a very high success rate. Such attackers might be neighbors, landlords, commercial data providers, or foreign agents. Especially for these unique individuals, if attackers find a match in the reconstructed data, they can be highly confident in the accuracy of their reidentification of that individual's race and ethnicity.  And if they reconstructed additional microdata (e.g., relationship to householder, number of residents, number of children, racial composition of the household), those would also be accurate for the same reason---the person remains a population unique on block, sex and age. 

\begin{table}[h]
\label{table:reid:uniques}
\tabcolsep7.5pt
\caption{Reidentification rates for population uniques}
\begin{center}
\begin{tabular}{@{}l|l|c|c|c@{}}
\hline
Match file & Universe & Putative rate$^{\rm a}$ & Confirmed rate$^{\rm b}$ & Precision$^{\rm c}$\\
\hline
Commercial & All data defined persons$^{\rm d}$ &60.2\% &24.8\% &41.2\%\\
& Population uniques$^{\rm e}$ &23.1\% &21.8\% &94.6\%\\
\hline
CEF & All data defined persons$^{\rm d}$ &97.0\% &75.5\% &77.8\%\\
& Population uniques$^{\rm e}$ &93.1\% &87.2\% &93.6\%\\
\hline
\end{tabular}
\end{center}
\begin{tabnote}
\begin{flushleft}$^{\rm a}$The number of records that agree on block, sex, and age (exact/binned), divided by the total number of records in the universe. $^{\rm b}$The number of records that agree on PIK (the Census Bureau's internal person identifier), block, sex, age (exact/binned), race, and ethnicity, divided by the total number of records in the universe. $^{\rm c}$The number of confirmed reidentifications [records that agree on PIK, block, sex, age (exact/binned), race, and ethnicity] divided by the number of putative reidentifications [records that agree on block, sex, and age (exact/binned)].  $^{\rm d}$All individuals with a unique PIK identifier within the block (276 million persons for the 2010 Census). $^{\rm e}$All data defined individuals who are unique in their block on sex and exact/binned age. DRB clearance number CBDRB-FY22-DSEP-004; Data are from J.M. Abowd et al. (manuscript under review) released in Hawes (2022). Abbreviations: CEF, Census Edited File; DRB, Disclosure Review Board; PIK, Protected Identification Key.
\end{flushleft}
\end{tabnote}
\end{table}


%


This level of precision for population uniques within a simulated attack is concerning  because it demonstrates that attackers could be very confident that they have accurately identified the individuals' race and ethnicity. Because individual blocks are often homogeneous, it may be relatively easy to infer an individual's race, independent of a reconstruction-abetted reidentification attack, by knowing the modal race/ethnicity of the block in which the individual resides, a statistic that is also not safe to release without applying SDL. Essentially, a simple classifier assigns the modal race/ethnicity to everyone in the block. This classifier has expected precision given by the ratio of individuals in the modal race/ethnicity to total persons in the block. Now consider the expected precision of the simple classifier in comparison to the precision possible using the reconstructed data for reidentification---in other words, the additional information learned about the individual through the leakage from releasing data beyond the modal race in the block. The precision rates for population uniques (on sex-age-block) vary depending upon whether the individual is in the modal or nonmodal race for that block. The expected precision for the simple modal-race classifier is 1 for population uniques who are in the modal race/ethnicity group and 0 for those in nonmodal race/ethnicity groups.  As one can see in Table 3, the precision rates for reidentifications are greater for population uniques of the modal race, but below the simple classifier rate of 1. However, the precision rates for reidentifications of population uniques of the nonmodal race are very high---well above 0.5 and much greater than the simple classifier precision of 0. This means that if attackers get a match for one of these individuals, they can be highly confident in the accuracy of the match, and their ability to infer the individual's race is entirely due to leakage of information from the individual's confidential census response.

\begin{table}[h]
\label{table:reid:modal}
\tabcolsep7.5pt
\caption{Reidentification rates for population uniques of the block's modal and nonmodal races}
\begin{center}
\begin{tabular}{@{}l|l|c|c|c@{}}
\hline
Match file & Population uniques$^{\rm a}$ & Putative rate & Confirmed rate & Precision\\
\hline
Commercial &All population uniques &23.1\% &21.8\% &94.6\%\\
&Of the modal race &25.3\% &24.2\% &95.3\%\\
&Of the nonmodal races &13.7\% &12.2\% &89.2\%\\
\hline
CEF &All population uniques &93.1\% &87.2\% &93.6\%\\
&Of the modal race &94.8\% &91.3\% &96.3\%\\
&Of the nonmodal races &86.2\% &70.2\% &81.5\%\\
\hline
\end{tabular}
\end{center}
\begin{tabnote}
\begin{flushleft}
$^{\rm a}$Individuals who are unique in their block on sex and exact/binned age. DRB clearance number CBDRB-FY22-DSEP-004. Data are from J.M. Abowd et al. (manuscript under review), released in Hawes (2022). Abbreviations: CEF, Census Edited File; DRB, Disclosure Review Board.
\end{flushleft}
\end{tabnote}
\end{table}

\section{STATISTICAL AND NONSTATISTICAL USES OF TABULAR DATA}
\label{sec:tabulations}

When they were produced for larger geographic areas and places with substantial populations, census data were primarily descriptive statistics that supported demographic analysis and regional economic planning models. As increasing amounts of statistical data were published for very-low-population geographic areas, the possibility of combining those tabular summaries with person- and household-level data to make inferences specific to individuals or households emerged. For example, data users began asking ``Is this person in my database a non-Hispanic white adult?'' or ``How many children are in this household in my school district database?'' Some inferences like this are statistical; others are nonstatistical as defined by CIPSEA, which codified Statistical Policy Directive 1 (OMB 2014) in statute. In the law’s definition, statistical purpose ‘‘(A) means the description, estimation, or analysis of the characteristics of groups, without identifying the individuals or organizations that comprise such groups; and (B) includes the development, implementation, or maintenance of methods, technical or administrative procedures, or information resources that support the purposes described in subparagraph (A)'' [44 U.S.C. § 3561(12) `Statistical Purpose'].

Distinguishing between statistical and nonstatistical inferences has been studied in SDL and in the computer science literature surrounding DP. In the SDL literature, the original definitions are provided by Dalenius (1977).
Dwork \& Naor (2010, p. 93) paraphrase Dalenius's definition of a \textit{statistical disclosure} as ``access to a statistical database should not enable one to learn anything about an individual that could not be learned without access,'' then provide a precise mathematical formulation.  

Dalenius meant the same thing by statistical disclosure as modern SDL claims. Consider his exact formulation in his original notation. He defined a universe population, $\{O\}_T$, frame, $\{O\}_F$, and population not in the frame, $\{O\}_{F^*}$ with $\{O\}_F \cap \{O\}_{{F^*}} = \emptyset$ and $\{O\}_T = \{O\}_F \cup \{O\}_{F^*}$. Each element of $\{O\}_F$ is given by $(I_k, C_k)$, where $I_k$ is an identifier and $C_k$ is a vector of characteristics with elements $(X_k, Y_k, \dots,Z_k)$. A set of statistics, $S$, is the output of a function, $M$, that maps $\{O\}_F$ into $\mathcal{Z}$. The formal definition is: $M: \{O\}_F \rightarrow \mathcal{Z}$, $\mathcal{Z}$ countably infinite, and $S \subset \mathcal{Z}$ Lebesgue measurable. Finally, he defined external information $E_k$ that contains data about $I_k$ that are separate from the statistical database but may have elements in common with  $C_k$---what we would now call the attacker's information. Now consider an element of the population, $O_k \in \{O\}_T$, either $O_k \in \{O\}_F$ or $O_k \in \{O\}_{F^*}$. Consider a characteristic, $D_k$, where $D_k \in C_k \cup E_k$. Dalenius's (1973, p. 433) exact words are ``if the release of the statistics $S$ makes it possible to determine the value $D_k$ more accurately than is possible without access to $S$, a disclosure has taken place; more exactly, a $D$-disclosure has taken place.'' Dwork \& Naor (2010), and many others, quantify this definition using prior-posterior learning. It can be met if
\begin{equation}
    \label{eq:prior_posterior}
    \frac{P[D_k=d|M(\{O\}_F) \in S,E]}{P[D_k=d|E]} \leq \exp{(\epsilon)},
\end{equation}
where $\exp({\epsilon})$ is bounded for all $S$ and $E$---that is, if what is now called the inferential disclosure can be bounded. Dwork \& Naor (2010) prove that it cannot be bounded for all attacker information sets, although bounds may exist for specific attacker information sets. More importantly, they show that the inferential disclosure problem is incompletely specified. It is imprecise about the hypotheses under consideration, failing to distinguish statistical and nonstatistical inferences in the CIPSEA sense.

Many parables have been told to explain why Equation (\ref{eq:prior_posterior}) is an incorrect specification of a statistical disclosure. Dwork \& Pottenger (2013) provide a compelling example based on the work of Hammond \& Horn (1954), who were the first to publish statistics linking cigarette smoking to lung cancer death [per the American Cancer Society (\url{https://www.cancer.org/research/population-science/cancer-prevention-and-survivorship-research-team/acs-cancer-prevention-studies/history-cancer-prevention-study.html})]. Their example concerns the difference between the statistical inference that a smoker is more likely to have lung cancer than a nonsmoker versus the prohibited nonstatistical inference that a particular person has lung cancer because that person provided a data record to a survey with the contents $(I_k=millennial,X_k=smoker,Y_k=lung\_cancer)$, where $millennial$ is the identifier of person $k$ who is a millennial (born $1981-1996)$, $smoker_k$ is a binary variable indicating if person $k$ is a smoker, and $lung\_cancer_k$ is a binary variable indicating if person $k$ has lung cancer. In Dalenius's framework, whether one smokes is an element of $E_k$, the external or attacker data, and $C_k$, the data collected by the agency. Whether one has lung cancer is only an element of $C_k$. The statistic $S$, the output of the function $M$ in Equation \ref{eq:prior_posterior}, is the ratio of smokers who have lung cancer to total smokers in the frame $\{O\}_F$ (assuming no sampling, as Dalenius did, for simplicity). In this study, $\{O\}_T$ is adult males living in the United States; $\{O\}_F$ is the 187,866 males in the original cohort in 1952; and $\{O\}_{F^*}$ is all other males living in the United States, in particular, those born after 1952. We observe a millennial male smoking today. We properly conclude that this person is more likely to die of lung cancer. So do his life and health insurance companies. The statistic $S$ is the basis for this inference, but the millennial is clearly in $\{O\}_{F^*}$. So, $S$ cannot be a confidentiality breach. The record  $(I_k=millennial,X_k=smoker,Y_k=lung\_cancer)$ cannot be in $\{O\}_F$.  It must be in $\{O\}_{F^*}$. Even though the inference that the smoking millennial is more likely to die from lung cancer and the associated insurance premium penalties are clearly harmful, they are not a statistical disclosure because they cannot have relied on private information provided by the participant to the 1954 study, as the millennial was not even alive at the time of the study. Limiting the difference between prior and posterior also means that the causal inference ``smoking causes cancer'' must be suppressed in the released data. In other words, using Equation (1) to limit inferential disclosure necessarily also prevents the disclosure of generalizable statistical knowledge---an outcome that would severely hamper statistical agencies' mandate to produce reliable, trustworthy data.

The DP framework (Bun \& Steinke 2016; Dong et al. 2020; Dwork 2006; Dwork \& Rothblum 2016; Dwork et al. 2006a,b; Mironov 2017) and the posterior-to-posterior inference system it supports (Kasiviswanathan \& Smith 2014) resolve the Dalenius paradox. We will stay in his notation. The frame without individual $k$ is 
$\{O\}_F^{-k} = \{\{O\}_F \setminus \{(I_k,C_k)\}\}$. Now consider
\begin{equation}
    \label{eq:posterior_posterior}
    \left. \left[\frac{P[D_k=d|M(\{O\}_F) \in S,E]}{P[D_k=d|E]}\right] \middle/
    \left[\frac{P[D_k=d|M(\{O\}_F^{-k}) \in S,E]}{P[D_k=d|E]}\right] \right. \leq \exp{(\epsilon)},
\end{equation}
which is the ratio of the posterior distribution from Equation (\ref{eq:prior_posterior}) when entity $k \in \{O\}_F$ to the posterior distribution when entity $k \not \in \{O\}_F$. Notice that the prior distributions cancel, and that the released statistics $S$ have the same value in the numerator and denominator. Equation (\ref{eq:posterior_posterior}) measures the influence of the record $(I_k=millennial,X_k=smoker,Y_k=lung\_cancer)$ when the released statistics are $S$, and DP frameworks provably bound it. It quantifies how important individual $k$'s data were to any inference based on the statistic $S$. Wasserman \& Zhou (2010, theorem 2.4, p. 377) used this argument to show that all tests of the hypotheses $H_0:\{O\}_F$ versus $H_1:\{O\}_F^{-k}$ based on Equation (\ref{eq:posterior_posterior}) have power, $(1-\beta)$, bounded by $(1-\beta) \leq \exp({\epsilon}) \ell$, where $\ell$ is the level and $\beta$ is the probability of a type II error of the test, respectively. 

Textbook definitions are similarly ambiguous. An \textit{identity disclosure} occurs when ``a data subject is identified from released data'' (Duncan et al. 2011, p. 174). An \textit{attribute disclosure} occurs when ``information [is disclosed] about a population unit without (necessarily) the identification of the unit within the data set'' (Duncan et al. 2011 p. 172). One must distinguish between identity and attribute inferences that depend upon the use of the protected entity's data and those that are possible without using the protected entity's information. The confidentiality-breaching versions of identity and attribute inferences are special cases of the problem summarized in Equation (\ref{eq:posterior_posterior}). An identity disclosure is the hypotheses $H_0:(I_k,C_k) \in \{O\}_F$ versus $H_1:(I_k,C_k) \not \in \{O\}_F$; that is, using Equation  (\ref{eq:posterior_posterior}), one learns $I_k$ is associated with the characteristics $C_k$. An attribute disclosure is the hypotheses $H_0:(I_k,X_k,Y_k, \dots, Z_k) \in \{O\}_F$ versus $H_1:(I_k,X_k,\tilde Y_k, \dots, Z_k) \in \{O\}_F$, for some arbitrary, prespecified $\tilde Y_k$; that is, already knowing the identity $I_k$, the attacker learns the attribute $Y_k$. Both are special cases of reidentifying inferential disclosure.

\section{GLOBAL MEASURES OF DISCLOSURE RISK}
\label{sec:accounting}

Effectively mitigating the risk of inferential disclosure in the context of multiple data releases derived from the same confidential source requires a mechanism for identifying and quantifying the potential for reidentifying inference across all data releases, while enabling statistical inferences in the CIPSEA sense. This global privacy-loss accounting is what separates DP frameworks from most traditional methods of disclosure limitation. DP frameworks properly distinguish between statistical inference and reidentifying inference. 
Importantly, they permit the data publisher to do consistent global risk analysis over a confidential data source's entire publication cycle. For the 2020 Census, this means over the 72 years from April 1, 2020, until the National Archives releases the confidential records on April 1, 2092. There are now several DP frameworks: pure DP (Dwork et al. 2006b), approximate DP (Dwork et al. 2006a), concentrated DP (Dwork \& Rothblum 2016), zero-concentrated DP (z-CDP) (Bun \& Steinke 2016), R\'enyi DP (Mironov 2017), and $f$-DP (Dong et al. 2020), among others. Each of these frameworks offers some advantages for certain statistical releases and disadvantages for others. Each is supported by a variety of DP mechanisms---randomized algorithms that implement a particular DP framework. Brenner \& Nissim (2014) proved the impossibility of a universally optimal DP mechanism. 

Privacy-loss accounting provides the essential tool for limiting reidentification inferences by quantifying the ex ante incremental risk of these inferences across all publications. This quantification is enabled by the composition property of DP. Because mechanisms satisfying DP frameworks compose, they can reliably quantify the incremental reidentification risk of a particular subset of the publications given the ensemble. The innovation of privacy-loss accounting is comparable to the invention of double-entry bookkeeping in accounting systems. It permits external verification of the strength of the confidentiality protection that is as independent as possible of arbitrary assumptions about the attacker's information, methods, and computational strength. Traditional SDL systems do not have this property. 

In principle, the privacy-loss accounting afforded by various DP frameworks is similar to the accounting used in conventional primary/complementary cell suppression (e.g., Willenborg \& de Waal 2001, p. 78; Data Protection Toolkit at \url{https://nces.ed.gov/fcsm/dpt/content/3-2-2-1}).  To use this SDL method, one specifies a \textit{primary suppression} rule that defines the conditions under which the frequency count or magnitude value in a particular cell cannot be published. This primary suppression rule is a policy decision; that is, there is no mathematical definition that lists properties a suppression rule must satisfy in order to ensure that the suppressions protect confidentiality. \textit{Complementary suppression} is the process used to suppress additional cells to ensure that the primary suppressed values cannot be recovered using a particular reconstruction method known as a \textit{subtraction} attack. Here, the mathematics are much more precise.  Cox (1995) uses linear optimization over a network to find the optimal complementary suppression from a given primary suppression rule, where optimality is defined as the minimum amount of complementary suppression (in an appropriate metric) required to ensure the primary suppressions cannot be recovered. If all the tables to be released from a given confidential input are included in the network model and if the resulting linear program can be solved, the primary/complementary suppression method that Cox proposed provably protects the values in the primary suppression cells. The Census Bureau implemented these methods for tables based on surveys and censuses of businesses (Massell 2001, Steel 2013).

While properly designed cell suppression systems enjoy some of the properties of DP frameworks, they have several major disadvantages. First, while they are often labeled \textit{nonperturbative}, meaning the published values in the unsuppressed cells equal the values tabulated from the confidential data, the published data exhibit nonignorable missingness (Little \& Rubin 2002, pp. 119-120 and chapter 15). Nonignorable missing data can generate serious biases when the published data are analyzed (Abowd \& Schmutte 2015). Second, suppression systems do not compose. If two different table sets are processed separately using the Cox network method, the resulting publications do not provably protect the primary suppressions in either set of tables. Uncoordinated use of cell suppression risks exposing some or all of the confidential data it was designed to protect. This is a particularly salient shortcoming because it limits the number of table sets that can be published from a confidential source in a path-dependent way. Programs that release data sequentially, such as the Economic Censuses conducted by the Census Bureau, are particularly vexed by this limitation. The linear programs they must solve become increasingly complex as additional tables are added to the set previously released to ensure that all primary suppressions are properly protected throughout the publication process.

\section{ALTERNATIVES}
\label{sec:alternatives}

Faced with compelling mathematical and empirical evidence of the inherent vulnerability of the 2010 Census swapping mechanism in protecting against reconstruction-abetted reidentification attacks, the Census Bureau began exploring alternative strategies that it could employ for the 2020 Census. The three primary methodologies the agency considered were suppression, enhanced data swapping, and DP.

While the Census Bureau could use suppression to mitigate disclosure risk from a reconstruction attack, most census data would be only available at a very high level of aggregation because of the necessary complementary suppressions. This was the reason suppression was abandoned after the 1980 Census (McKenna 2018, p. 11). Today’s data users, including redistricters, rely on detailed block- and tract-level data, which would not be available for many areas if the Census Bureau were to return to suppression to protect against modern attacks. Additionally, suppression rules disproportionately impact small population groups and small areas because of the nonignorable missing data problem (see Section \ref{sec:accounting}). 

Noise infusion via swapping had worked for three censuses. Enhancing the data swapping mechanism used for the 2010 Census in a manner sufficient to protect against emerging threats like reconstruction attacks would have a significant, detrimental impact on data quality. Thus, the focus was on using noise infusion through DP because it would provide the necessary protections and better-quality data. With an estimated 57\% of the population\footnote{57\% of the 308,745,538 person records in the confidential 2010 CEF, the definitive source for all 2010 Census tabulations, were unique on their block location, sex, age (in years), race (any combination of the 6 Office of Management and Budget-approved race categories, 63 possibilities in all) and Hispanic/Latino ethnicity. This previously confidential statistic was approved for publication with DRB clearance number CBDRB-FY21-DSEP-003.} known to be unique at the block level, a swapping mechanism that targets vulnerable households for swapping would require significantly higher rates of swapping than were used in 2010 to protect against a reconstruction attack (Hawes 2022). Implementing swapping in 2020 would also require abandoning the total population and voting age population invariants that were used in 2010. There are two technical reasons for this. First, at swap rates sufficient to counter the reconstruction of microdata accurate enough to enable large-scale reidentification, it is impossible to find enough paired households with the same number of persons and adults without searching well outside the neighborhood of the original household. Finding swap pairs was a challenge for some states even at the 2010 swap rate. Second, holding the total and adult populations invariant gives the attacker a huge reconstruction advantage—exact record counts in each block for persons and adults. This advantage vastly improves the accuracy of the reconstructed data. Even a small amount of uncertainty about the block location of an individual greatly expands the variability in the reconstructed microdata, effectively reducing the chances of a correct linkage in a reidentification attack (Hawes 2022). If a block is known to contain exactly seven persons in the confidential data, then every feasible reconstructed version of those data will have exactly seven records in that block, meaning that the block identifier will be correct on every record of every feasible reconstructed database. But if the block population is reported with some random fluctuation around seven, then only by chance will the block identifier be correct in the reconstructed data. Compound this effect over 5,892,698 blocks (the number with potential positive population in the 2020 Census) and the number of feasible reconstructions explodes exponentially. This is what provides the protection against reidentification from the reconstructed data (Garfinkel et al. 2019). 


DP offers a more efficient trade-off between confidentiality and accuracy. This can be seen empirically in the results of recent experiments. In these experiments, the Census Bureau generated a privacy-protected microdata file from the confidential 2010 Census data using an enhanced version of the 2010 swapping mechanism similar to the methodology described by Hawes \& Rodriguez (2021). They then compared the relative accuracy and disclosure risk of the enhanced swapped data to the recent Demonstration Data Product 2022-03-16, which was protected using DP. Table 4 provides a representative comparison of the resulting error from the two methods, and Table 5 shows the reidentification statistics for these same files. As these tables show, the enhanced swapping mechanism does reduce disclosure risk compared with the swapping mechanism used in 2010 for this specific attack, but at a significant cost to data accuracy. While the particular implementations of these competing mechanisms are based on policy choices---swapping rate, privacy-loss budget (PLB) allocation---that can be legitimately debated (how much protection is sufficient?), the empirical results of these experiments support the conclusion that for any given level of confidentiality protection, the targeted application of a DP mechanism can produce more accurate data than known non-DP solutions.

\begin{table}[h]
\label{table:accuracy}
\tabcolsep7.5pt
\caption{Accuracy of 2010 Census, enhanced Swap, and DP: mean absolute error (in persons) for age group population counts at the county level}
\begin{center}
\begin{tabular}{@{}r|c|c|r@{}}
\hline
Age group & 2010 Census & Enhanced swap & DP \\
\hline
0-17 years &0 &256.41 &9.84\\
18-64 years &NA$^{\rm a}$ &494.16 &12.83\\
65 years and over &NA$^{\rm a}$ &431.37 &12.66\\
\hline
\end{tabular}
\end{center}
\begin{tabnote}
\begin{flushleft}$^{\rm a}$Error statistics for the impact of swapping as applied to the published 2010 Census are confidential. The 2010 Census swapping algorithm kept the number of non-voting age individuals (0-17 years) invariant but did inject noise into the age groups within the voting age population. DRB clearance number CBDRB-FY22-DSEP-003. Data are from Devine \& Spence (2022). Abbreviations: DP, differential privacy; DRB, Disclosure Review Board; NA, not available.
\end{flushleft}
\end{tabnote}
\end{table}

\begin{table}[h]
\label{table:reid:swap}
\tabcolsep7.5pt
\caption{Reidentification statistics for 2010 Census, enhanced swap, and DP}
\begin{center}
\begin{tabular}{@{}r|c|c|c@{}}
\hline
Reidentification Statistic & 2010 Census & Enhanced swap & DP\\
\hline
Putative reidentification rate & 97.0\% &75.4\% &44.4\%\\
Confirmed reidentification rate & 75.5\% &46.6\% &27.4\%\\
Precision rate & 77.8\% &61.8\% &61.7\%\\
\hline
Precision for population uniques (nonmodal race) & 81.4\% &33.4\% &24.0\%\\
\hline
\end{tabular}
\end{center}
\begin{tabnote}
\begin{flushleft}DRB clearance number CBDRB-FY22-DSEP-004. Data are from J.M. Abowd et al. (manuscript under review), released in Hawes (2022). External Matching File: Census Edited File. Abbreviations: DP, differential privacy; DRB, Disclosure Review Board.
\end{flushleft}
\end{tabnote}
\end{table}

An alternative that was not considered was the enhanced use of synthetic data based on the models that were adopted for group quarters in the 2010 Census (McKenna 2018, p. 10). These methods use Rubin's (1993) and Little's (1993) definition of synthetic data as samples from the posterior distribution of the confidential data replacing some (Little) or all (Rubin) of the responses with these samples. To implement this approach would have required basic research and testing on a scale even greater than implementing the DP framework, and a prototype redistricting application could not have been developed in time for the 2018 End-to-End Census Test. The Census Bureau investigated this approach for the 2022 Economic Census (Kim et al. 2020, Thompson et al. 2020) because it shows promise in supporting a tiered-access system that automates the validation procedures now used for synthetic versions of the Survey of Income and Program Participation and Longitudinal Business Database (Benedetto et al. 2018, Kinney et al. 2011). The Census Bureau is also researching this approach for the American Community Survey (Rodriguez 2021).



\section{THE 2020 CENSUS DISCLOSURE AVOIDANCE SYSTEM}
\label{sec:DAS}

The suite of algorithms developed to protect the 2020 Census are known collectively as the 2020 Census Disclosure Avoidance System (DAS). A principal component of this system is the TopDown Algorithm (TDA), which was used to protect the redistricting data. The TDA uses the framework called z-CDP, implemented via the discrete Gaussian mechanism. The inputs are record-level confidential microdata and the final tabulation geography. The output is record-level microdata with the same schema as the input data for the tabulation variables. Most of the details are provided in a separate article (Abowd et al. 2022). This section covers only the essentials.

Planning for the systems that ingest, process, and tabulate decennial data begins early in the decade preceding the census. The agency's decision in 2017 that prior decades' SDL would be insufficient to protect the 2020 Census from known vectors of attack specified that the new SDL solutions would have to be implemented within an existing data processing ecosystem. For the 2000 and 2010 Censuses, similar decisions were made in 1999 (to extend swapping to the redistricting data) and 2008 (to implement enhanced swapping and synthetic group quarters data), respectively. The 2017 DSEP decision was directed to the 2018 End-to-End Census Test, which was required to publish sample redistricting data. As with all 52 systems used for the 2020 Census, the versions used in the end-to-end test were prototypes for the full decennial implementation.

Here are the main constraints. First, the published tables had to be internally and hierarchically consistent. 
Second, the privacy-protected output of the TDA had to be ingested by the existing tabulation system used to produce the redistricting data tables. These design requirements forced the DAS to output privacy-protected microdata without any negative or fractional values that other DP implementations might produce. DSEP also required certain tabulations to be published as enumerated, i.e., without any noise infusion. These invariants were the total resident population counts at the state level, the number of housing units at the block level, and the number and major type of occupied group quarters facilities at the block level. The overall PLB and the allocation of that PLB were controlled by DSEP to ensure that accuracy targets were met for specific tabulations at specific geographic levels. Lastly, the system met quality standards for official statistics---the relative accuracy of the estimates increases with the size of the population being measured.

The TDA ingests the variables in the CEF that are specified in the publication tables and the corresponding Geographic Reference File, then converts the CEF microdata into a functionally equivalent \textit{histogram}---the vectorized fully saturated contingency table with structural zeros removed or imposed by constraints. Next, the TDA makes a series of queries of the confidential data called \textit{noisy measurements}. The DAS uses z-CDP with privacy-loss parameter $\rho$ implemented using noisy measurements from the discrete Gaussian mechanism (Canonne et al. 2020, 2021). The parameter $\rho$ is used for privacy-loss accounting.  The TDA postprocesses the resulting noisy measurements to ensure internal and hierarchical consistency and to preserve the specified invariants and other editing constraints, resulting in a fully specified privacy-protected histogram at the block level that can be converted directly into privacy-protected microdata. The tabulation system processes these microdata into statistical tables for release.

A simple implementation of the DP framework might concentrate exclusively on block-level queries of the confidential data because they aggregate to any higher geographic level. This bottom-up approach, however, would result in privacy-preserving error that increases as one aggregates more and more blocks into larger geographies of interest, reducing the overall statistical utility of the resulting data (Abowd et al. 2022, appendix A). Instead, the DAS takes a top-down approach, which spreads the PLB strategically across geographic levels and queries to control the statistical accuracy better than the bottom-up approach could. The bottom-up approach is similar to the Census Bureau's first DP implementation, OnTheMap (Machanavajjhala et al. 2008, USCB 2022b), and the way technology companies like Google (Bittau et al. 2017), Apple (Apple Differ. Priv. Team 2017), Uber (Johnson et al. 2018), and Microsoft (Ding et al. 2017) use local DP, although bottom-up would still outperform local DP applied directly to the CEF. Technology companies use  local DP to prevent the ingestion of the confidential information in the first place; the DP mechanism acts on the data as they are ingested.

The TDA uses a custom geographic hierarchy that begins at the level of the entire United States and descends through intermediate geographic levels down to individual blocks (Abowd et al. 2021a), which are the atoms of the agency's geographic lattice.\footnote{Blocks are defined by the Geography Division at the Census Bureau precisely for this purpose, and not primarily to serve as statistical tabulation areas themselves (USCB 1994, chapter 11).} Each geolevel of the hierarchy is a partition of the United States (i.e., the set of geounits within each geolevel are mutually exclusive and exhaustive). The TDA starts with the noisy measurements of the US-level histogram along with the corresponding US-level invariants and constraints, and postprocesses those into an internally consistent privacy-protected histogram for the entire United States.  Then, at each subsequent level down the geographic hierarchy, the algorithm repeats the process but includes the resulting internally consistent, privacy-protected histogram from the prior geographic level as an additional set of constraints during the postprocessing step. 

This top-down approach offers a number of significant advantages over the simpler bottom-up strategy. First, it controls the error at each level rather than accumulating error by aggregating from individual blocks into larger geographic areas on the geographic hierarchy while ensuring that all such aggregations are consistent in the final product. This was the primary design goal and reflects the principle that the relative accuracy of statistical tabulations should improve as the number of entities contributing to the calculation increases. Early research showed that a bottom-up approach could not accomplish this (Abowd 2018; Abowd et al. 2022, appendix A). The top-down approach allocates PLB strategically such that data for individual blocks (where disclosure risk is often greatest) remain relatively noisy, while ensuring statistically reliable data when aggregating those noisy blocks into larger geographic areas. This feature is critical for the redistricting use case, which is discussed in more detail in Section \ref{sec:uses}. 

The top-down approach addresses the challenge of sparsity in census data. A simple cross-tabulation of demographic characteristics in the redistricting data (voting age $\times$ race $\times$ ethnicity) yields 252 unique combinations. The average census block with living quarters contains 56 people in 2020. Thus, a very large proportion of histogram cells at the block level are empty. A bottom-up approach to DP in the context of these small cells, even when using a large PLB (implying a tiny variance in the discrete Gaussian mechanism), would introduce substantial noise into the counts for many of the less common combinations of demographic characteristics, resulting in substantial distortion in the overall demographic makeup of the nation. The top-down approach effectively mitigates the problem of sparsity by constraining counts for lower geographic levels to the values determined for the higher-level geographies, thus propagating zeros down the hierarchy as constraints. In this manner, statistics at lower levels of geography benefit from PLB allocated to geographic levels higher on the hierarchy. 
The TDA's progression down a geographic hierarchy allows tuning of the allocation of PLB to meet specified, use-case-driven accuracy targets for particular types of geographic areas. For example, the redistricting data were tuned by customizing the TDA's geographic hierarchy to include American Indian and Alaska Native tribal areas (in the 37 states with these areas) directly on the hierarchy. In addition, the TDA redefined tabulation block groups (for TDA purposes only) so that minor civil divisions (in the 12 states for which minor civil divisions are important governmental units) and places (in the balance of the states) were closer to the hierarchy using the metric defined by Abowd et al. (2021a).   

For z-CDP,  privacy loss, $\rho_{gj}$, is allocated first to the geographic levels $g$, then to each query set $j$ at that level. The composition property implies that the global $\rho$ is the sum of the $\rho_{gj}$ allocations to each query set at each geography level, except when the geographic hierarchy optimization reallocates between the parent and child nodes. Balle et al. (2020) show that if the full PLB of $\rho$ is assigned to a single query, then for all $\alpha > 1$ the level-power trade-off on reidentification inferences as defined in Section \ref{sec:changed}, is 
    $\ell^\alpha (1-\beta)^{1-\alpha} + (1-\ell)^\alpha \beta^{1-\alpha}
    \leq \exp(\rho\alpha(\alpha-1))$,
which is a quantified global reidentification risk that is impossible using swapping, suppression, or any other conventional SDL method.

The TDA parameters for the published redistricting data were primarily policy driven. The share of the overall 2020 Census PLB allocated to the redistricting data, and the allocation of that PLB to the individual queries taken of the confidential data, were set by DSEP. In setting these parameters, the agency had to consider and balance its countervailing obligations to produce high-quality statistics while also protecting the confidentiality of census respondents' data. In comparison with legacy SDL methods, the privacy-loss accounting framework of DP made evaluation of the overall strength of confidentiality protection relatively straightforward. Assessing and ensuring data quality and fitness for use of the resulting data for myriad important but diverse uses of census data, however, were more challenging. Accomplishing this required extensive engagement with a diverse array of internal subject matter experts, federal agency partners, and external data users and careful evaluation of the impacts of different parameter settings on the types of analyses and use cases on which these diverse stakeholders rely.      

\section{TARGETED USE CASES}
\label{sec:uses}

Article I, Section 2 of the US Constitution provides for a regular count of the population with its mandate that ``Representatives and direct Taxes shall be apportioned among the several States which may be included within this Union, according to their respective Numbers $\dots$ The actual Enumeration shall be made within three Years after the first Meeting of the Congress of the United States, and within every subsequent Term of ten Years, in such Manner as they shall by Law direct.'' 
Central to this requirement is the idea that seats in the House of Representatives (and, by extension, votes in the Electoral College) should be reallocated to each state every ten years based on their respective state-level population counts in the census, a process known officially as apportionment. 

Once the number of Congressional seats has been determined, however, census data serve another critical purpose in helping each state divide up its population geographically into the districts that elect representatives to fill those seats (\textit{redistricting}). State and local jurisdictions similarly rely on these population counts for the regular redrawing of legislative and voting districts for everything from state legislatures to city councils and school boards. Data beyond population counts are important to these redistricting activities. Civil rights laws, most notably the VRA, place additional importance on demographic characteristics (race, ethnicity, and voting age) to ensure that racial minorities and Hispanic ethnicity populations receive fair treatment in the Congressional and voting district maps that get drawn. The importance of a regular census for the governance of the nation also goes far beyond the data's role in elections. James Madison, the so-called father of the Constitution, once said while debating the Census Act of 1790, ``in order to accommodate our laws to the real situation of our constituents, we ought to be acquainted with that situation...[the census] may ascertain it so far as to be extremely useful, when we come to pass laws, affecting any particular description of people'' (US House 1790). As such, one can easily say that among the most important use cases for the US Census are the apportionment of the House of Representatives, providing data in support of the redistricting of every legislative body in the country, and providing basic demographic inputs for the Population Estimates Program because many federal funding formulas and programs depend on the official population counts from the census and the Population Estimates Program.

Redistricting illustrates the fundamental tension between confidentiality protection and useful data dissemination. To draw new voting district maps, a minimum atom or pixel of geography must be prespecified as the mutually exclusive and exhaustive partition of the nation from which all districts will be composed. The partition must respect all political boundaries relevant to the legislative bodies that require redistricting. The atom must be small enough in population to permit one-person, one-vote districts as defined in law, regulation, and judicial precedent. Finally, when the atoms are composed into tentative districts, data on certain population characteristics must be sufficiently accurate to permit reliable enforcement of antidiscrimination laws. For all censuses since 1990, this atom was a census block and the required characteristics were child/adult, race, and ethnicity, with race and ethnicity collected and encoded in compliance with Statistical Policy Directive 15 (adopted in 1977, published as OMB 1978, revised as OMB 1997). While the 2020 Census confidentiality protections were being designed and implemented, the list was expanded to include citizenship, although it was later deleted. 

Many factors guide the redistricting process, and rules for drawing Congressional districts vary across states. That said, there are two principal federal requirements that all redistricting maps are required to meet. The first, deriving from the Constitution, is the equality standard of \textit{one person, one vote} that requires that all Congressional districts within a state have, to the greatest extent practicable, equal population counts (Congr. Res. Serv. 2017). The second, stemming from the 14th Amendment and Section 2 of the VRA, is the principle of \textit{equal protection}: States may not draw their Congressional districts in such a way as to ``have the effect of diminishing the ability of any citizens of the United States on account of race or color...to elect their preferred candidates of choice'' [52 U.S.C. § 10304(b)]. In practice, this latter requirement is often reflected in the imperative to draw majority-minority districts whenever possible. Claims of vote dilution in violation of the equal protection principle are evaluated using demographic data to determine if ``the minority group$\dots$is sufficiently large and geographically compact to constitute a majority in a single-member district'' under the US Supreme Court's three-pronged Gingles Test (\textit{Thornburg v. Gingles} 1986).


To meet these requirements, those tasked with redrawing districts need data on population counts and the racial and ethnic composition of geographic units of sufficient granularity that those geographic units can be aggregated into geographies of equal population size with, when appropriate, sufficiently sized populations of a particular racial or ethnic minority to avoid claims of vote dilution under the Gingles Test. In practice, the geographic atom used to build these maps is the individual census block.

Most official statistics are published for standardized tabulation geographies. For Census Bureau products, these follow the standard tabulation geographic hierarchy of nation, state, county, tract, block group, and block (see USCB 2021a, figure 2-1). Other geographies of interest, such as incorporated places, American Indian reservations (see USCB 2021a, figure 2-2), or school districts that do not fit neatly within this hierarchy are tabulated from the data for the lower-level geographic areas that they contain. When selecting and implementing any disclosure limitation strategy, producers of official statistics are aware that the implementation of any methodology will impact the accuracy or fitness for use of the resulting publication in differing ways. They evaluate these impacts on data for geographies of interest against known use cases and adjust their disclosure limitation methods accordingly if the impact on data quality is deemed too great.



The fundamental challenge with the redistricting use case is that the geography of interest, the Congressional or voting district to be drawn, cannot be known prior to the publication of the data used to draw it. This precludes the possibility of tuning the SDL to the specific geographic areas that will be drawn during redistricting. Instead, the disclosure limitation needs to find a way to ensure sufficient protection of the block-level data, where disclosure risk is greatest, while also ensuring that when aggregating geographically contiguous blocks into Congressional or voting districts, the resulting data are sufficiently accurate to evaluate the newly drawn districts for adherence to the equality and equal protection standards central to the redistricting process. 

The known error distributions of DP mechanisms make this process somewhat easier compared with swapping.  
The relative accuracy of statistics can be estimated for any geographic entity on the TDA's geographic hierarchy. 
Properties of the DP mechanisms can be used to tune the algorithm's parameters to ensure specific accuracy targets are met for these geographies. For geographic areas off of the hierarchy, error can be estimated (and thus tuned) based on the proximity or distance of the geography from the central hierarchy, i.e., the minimum number of on-hierarchy geographic areas that must be added or subtracted to construct the off-hierarchy geographic area (Abowd et al. 2021a, p. 11). 


To ensure that the DAS was appropriately tuned to meet the needs of data users, the Census Bureau produced demonstration data generated from 2010 Census data, running them through the 2020 DAS with the same parameter settings as those used to protect the 2020 Census redistricting data. External stakeholders were invited to evaluate these data and provide feedback. Because external data users could only compare these data to published data affected by previous SDL procedures, internal teams also evaluated the data both for their own needs and by replicating the evaluations of external data users. For example, Wright \& Irimata (2021) performed an evaluation of the fitness for use of these demonstration data for the redistricting and VRA use cases.

Working with both internal and external data users, the Census Bureau was able to define an extensive set of accuracy targets. The agency conducted empirical analyses from over 600 full-scale runs of the DAS using 2010 Census data to determine the parameters necessary to ensure that the accuracy targets were met. Mean absolute error in total population counts for minor civil divisions and incorporated places in these demonstration data were 2.74 persons and 3.55 persons, respectively (USCB 2021b). For the very smallest places and minor civil divisions, those with total populations of 0-249, this error is even smaller (1.5 persons, with an approximate 90\% confidence interval of -3 to +3 persons) (USCB 2022a, table 3). This SDL-induced error is trivial when compared with other known sources of error in census statistics, such as coverage error and operational error (USCB 2022a, pp. 4-6).  

Tuning of the TDA to ensure fitness for use of the 2020 Census redistricting data, while effectively protecting confidentiality, was a major undertaking. That said, the redistricting data represent only a small portion of the tabulations produced from the census.  Population estimates and household statistics represent another major class of use cases.

Throughout the process of developing and optimizing the TDA, ongoing engagement with external data users was essential. From July 2018, when the Census Bureau released the first Federal Register Notice soliciting feedback, through publication of the redistricting data in August 2021, the Census Bureau received over 1,200 public comments. Each of these comments was carefully reviewed to inform the design and implementation of the TDA. Examples of pivotal feedback include statutory and programmatic use cases, measures of reliability for VRA analyses, diversity indices used by state and local demographers, indicators of postprocessing bias, and targets to ensure the accuracy of downstream use cases. Each of these helped inform subsequent tuning of the algorithm, and the major use cases and other accuracy measures were incorporated into the accuracy metrics the agency released with each set of demonstration data to report on its progress.

\section{SCIENTIFIC INFERENCE USING DIFFERENTIAL PRIVACY}
\label{sec:inference}

Traditionally, users of decennial census tabulations have ignored the uncertainty in those estimates. The only significant exception is the uncertainty caused by coverage error. Therefore, most direct uses of census tabulations do not account for the uncertainty due to operations, missing data, or SDL. Because the leading use cases---apportionment, redistricting, and population estimates---do not account for census uncertainty, the DAS was designed to minimize the effects of SDL uncertainty on key statistics. This decision affected the confidentiality protection. However, the uncertainty due to the DAS can be quantified without compromising the confidentiality protection. This section explains how to do so.

Because of the requirement to produce microdata, the error in any statistics produced by the DAS has two sources: the variability from the noisy measurements, which is fully characterized by the $\rho_{gj}$ allocation to the queries used to compute the statistic, and the postprocessing error from transforming the noisy measurements to microdata. The variance of each noisy measurement can be computed from the privacy-loss allocation tables (Abowd et al. 2022, tables 6-8); the global PLBs for the person tables, $\rho=2.56$ and the housing unit tables $\rho=0.07$ (Abowd et al. 2022, section 8.2); and the conversion formulas (Abowd et al. 2022, appendix B). This is all the information required to compute margins of error before post-processing, and it will produce reliable estimates for statistics calculated for entities on the TDA geographic hierarchy. 

Accounting for the postprocessing error is more difficult and is only necessary because of the microdata requirement. If the TDA were used to produce fully consistent tabular summaries with the same schema as the released redistricting tables, but without the inequality constraints required to generate microdata, there would be no postprocessing error. The formulas referenced above and a tractable way to compute the variances in equality-constrained weighted least squares are all that is required to do inference using these data. 
Abowd et al. (2021b) studied the problem of generating microdata via postprocessing, and showed that there exists a trade-off between accuracy on a marginal total and accuracy on its component cells. In this special setting, they established lower and upper bounds on the expected error that must be introduced by any postprocessing method for generating microdata. To extend their work to a quantitative method for adjusting statistical inference in the full TDA setting likely requires use of simulation methods.
As we discussed in Section \ref{sec:uses}, simulating this disclosure-avoidance-induced error by comparing the demonstration data products to the released 2010 Census data using variability across the same geounit provided reliable guidance that was confirmed when computationally expensive simulation methods were used (Abowd et al. 2022, Wright \& Irimata 2021). Ashmead (2019) proposed, and JASON (2020, p. 96) reviewed, a parametric bootstrapping procedure that also estimates the total variability due to the DAS.  

\section{LESSONS FOR THE FUTURE}
\label{sec:future} 

Statistical agencies cannot continue to expand both the amount and the quality of the data they provide from traditional sources without updating the methods they are using to protect confidentiality in the face of ever-increasing disclosure risk. Even with better SDL methods, some reexamination of the quantity of published statistics will likely be necessary. 

The Census Bureau's modernization of SDL for the 2020 Census marks a pivotal moment in the history of official statistics (Meng 2020, Sullivan 2020). Parallel advancements in computing power and the availability of detailed external data sources have rendered traditional SDL methods largely insufficient to protect confidentiality at scale. But the agency's transition to the DP accounting framework for the 2020 Census also marks a turning point in the way that national statistics offices and statistical agencies across the federal government can and should begin assessing their policies and tools for protecting the confidentiality of their data subjects' information. Not all statistical data products carry the same inherent level of disclosure risk as the highly granular US decennial census, so many agencies may opt for incremental improvements to their traditional confidentiality protection mechanisms for the time being rather than wholesale abandonment of these legacy methods. The Census Bureau itself has taken this incremental approach for many of its surveys while it researches and assesses how it might apply formal privacy methods in a manner that ensures the continued high quality of those products. The agency's own research demonstrated, however, that for the 2020 Census, modernization was necessary and urgent. The experience, and  missteps, in this modernization process offer a number of useful lessons that may benefit others considering  taking this path.  Key lessons for the future include the need to: (\textit{a}) acknowledge at the outset the limitations for publication caused by the increasing difficulty of protecting confidentiality, (\textit{b}) consult and engage with data users and key stakeholders early and often throughout the implementation process, and (\textit{c}) design the publication system and the SDL simultaneously instead of requiring the SDL to work inside an existing publication system. 

SDL practitioners have long acknowledged that it is fundamentally impossible to completely eliminate disclosure risk in any release based on a confidential data source (Cox 1976, Fellegi 1972, OFSPS 1978). A statistical agency's duty is to manage this risk responsibly. The spectre of successful database reconstruction at scale confirms that in statistical products, data accuracy and data protection are finite resources that must be carefully managed (Abowd \& Schmutte 2019), especially as sophisticated attacks on statistical data products have become ever easier due to advances in computing power and continued improvements in external data sources. Historically, however, the general trend in official statistics has been the expansion of data products to include new geographies, new variables, new disaggregations, and so on, largely driven by rising data-user demand for more granular and timely data, increasing reliance on statistical data for program administration and evidence-building, and the development of new and better techniques for ingesting and processing data from nontraditional sources. These two realities are in stark opposition. Formal privacy-loss accounting frameworks offer a number of solutions for addressing this challenge, but effective adoption of this approach requires first acknowledging these competing interests. In the context of the 2020 Census, this meant accepting the fact that protecting against unlawful disclosure required eliminating some of the tables that data users wanted.

Perhaps one of the greatest lessons learned from the Census Bureau's experience transitioning to DP is the active role that data users can and should play in shaping the disclosure limitation system. The SDL method an agency selects and the manner in which the SDL is implemented can have profound consequences for the quality of the data produced. Different methods impact the data differently, and choices and trade-offs always have to be made. These decisions have significant implications for certain data use cases. Historically, however, most statistical agencies have made these decisions internally, with little to no transparency to the data user community.

The Census Bureau discovered early in the transition to DP that it had done a poor job of communicating with the broader data user community that noise had been injected into prior censuses to protect confidentiality. The agency also discovered that its data users did not have the information necessary to understand how disclosure limitation noise intersected with other types of error and data limitations. Statistical agencies must develop tools that more actively communicate statistical uncertainty.

Transparency alone is not enough. Governing a risk accounting system requires making trade-offs that impact data users. Statistical agencies around the world employ motivated, talented statisticians, demographers, economists, and other experts. But, as knowledgeable as these experts are, they rarely know the myriad uses to which the public data user community puts their data products. Statistical agencies need to work with the public to better understand their needs.

The Census Bureau benefited enormously from its engagement with data users and other stakeholders as it was tuning and optimizing the TDA algorithm for the production of the redistricting data. Crowd-sourcing analysis of the demonstration data allowed the agency to develop a much more thorough understanding of where accuracy needs were not being met than they would have from an internal analysis alone. That said, the Census Bureau's external stakeholders were justifiably disappointed and concerned about the timing, pace, and mechanisms of the agency's public engagement regarding the development of the 2020 DAS. As discussed above, the urgency of the need to transition to a new disclosure avoidance mechanism both precluded some stakeholder engagement that would have been valuable and condensed the time-frame for the engagement that could occur. Going forward, however, the Census Bureau must explore additional approaches to engaging stakeholders. Transparency in the decision-making process regarding disclosure limitation is incredibly valuable, as it can help data users better understand any relevant limitations of the data they are using. However, engaging with data users and other stakeholders as part of that decision-making process is also necessary to ensure that data are fit for use (JASON 2020). 

Because DP frameworks are privacy-loss accounting frameworks, rather than specific SDL techniques, they are remarkably versatile in implementation options. That said, some implementations will be more efficient at managing the overall privacy/accuracy trade-off or at producing useful statistics than others. Using existing publication design requirements for a DP system based on predetermined data product specifications or existing data processing infrastructure can significantly reduce that versatility and may necessitate making some difficult compromises. Because the Census Bureau modernized disclosure avoidance in a modular fashion for the 2020 Census, retaining systems that created the CEF and produced the publication tables, there were design constraints imposed on the DAS independent of the DP framework, including specifications for the content and structure of the redistricting data and the need for the system to output privacy-protected microdata. 

These constraints posed a number of challenges that were difficult to address. Post-processing to generate microdata output introduced an upward bias for small populations and a corresponding downward bias for large ones (Comm. Natl. Stat. 2020, Hawes 2020). This result is due to the inequality constraints required to produce microdata (Abowd et al. 2021b) and not the DP framework. While improvements in algorithm design and careful tuning of the query structure and PLB allocations were able to effectively mitigate these distortions to relatively negligible levels compared with other sources of error in the census, they could not be eliminated entirely (USCB 2021b). These design constraints also prevented external privacy experts from evaluating the data effectively. Without those design constraints, more effective solutions would have been possible.

Statistical agencies like the Census Bureau are tasked with collecting and publishing data. To achieve these goals, they must guarantee confidentiality to the public (especially those who are most vulnerable) while also ensuring that published data are statistically sound. DP offers a valuable technical mechanism for governing these competing commitments, but a successful implementation also requires attending to how the data is understood by data users and the public.

\section*{DISCLOSURE STATEMENT}
The authors are not aware of any affiliations, memberships, funding, or financial holdings that might be perceived as affecting the objectivity of this review. 

\section*{ACKNOWLEDGMENTS}
The views expressed in this article are those of the authors and not the US Census Bureau. We thank Karen Battle, danah boyd, Ryan Cumings-Menon, Jason Devine, Michele Hedrick, Daniel Kifer, Alexandra Krause, Philip Leclerc, Jennifer Shopkorn, and Kathleen Styles. This article is forthcoming in the \textit{Annual Review of Statistics and Its Applications} (2023). This preprint was prepared on December 27, 2022 to make the preprint conform to the final journal copy edits. Preprint versions contain verified links for all cited work. If there are additional pre-publication edits, the arxiv version [\url{http://arxiv.org/abs/2206.03524}] will be updated.



\section*{LITERATURE CITED}

\noindent 
Abowd J. 2017. 
How will statistical agencies operate when all data are private? 
\textit{J. Priv. Conf.} 7(3):1. 
\url{https://doi.org/10.29012/jpc.v7i3.404}

\vspace{3mm}\noindent 
Abowd J. 2018. 
\textit{The US Census Bureau adopts differential privacy.} Invited lecture at the 24th ACM SIGKDD International Conference on Knowledge Discovery \& Data Mining, London, Aug. 19-23. 
\url{https://dl.acm.org/doi/10.1145/3219819.3226070}

\vspace{3mm}\noindent 
Abowd J. 2021a. 
Official statistics at the crossroads: data quality and access in an era of heightened privacy tisk. 
\textit{Surv. Statist.} 83:23--26.  \url{http://isi-iass.org/home/wp-content/uploads/Survey_Statistician_2021_January_N83_03.pdf} cited on April 6, 2022

\vspace{3mm}\noindent 
Abowd J. 2021b. 
Second declaration of John M. Abowd, appendix B: 2010 reconstruction-abetted re-identification simulated attack. \textit{Fair Lines America Foundation, Inc. v. U.S. Department of Commerce}.  \url{https://www2.census.gov/about/policies/foia/records/disclosure-avoidance/appendix-b-summary-of-simulated-reconstruction-abetted-re-identification-attack.pdf} cited on April 6, 2022

\vspace{3mm}\noindent 
Abowd J, Ashmead R, Cumings-Menon R, Kifer D, Leclerc P, Ocker J, Ratcliffe M, Zhuravlev P. 2021a.
Geographic spines in the 2020 Census Disclosure Avoidance System TopDown Algorithm. 
\url{https://arxiv.org/abs/2203.16654} [cs.CR]

\vspace{3mm}\noindent 
Abowd J, Ashmead R, Cumings-Menon R, Garfinkel S, Kifer D, Leclerc P, Sexton W, Simpson A, Task C, Zhuravlev P.
2021b. 
An uncertainty principle is a price of privacy-preserving microdata. In \textit{35th Conference on Neural Information Processing Systems (NeurIPS 2021)} ed. M Ranzato, A Beygelzimer, Y Dauphin, PS Liang, J Wortman Vaughan, pp. 11883-95. N.p.:NeurIPS.
\url{https://proceedings.neurips.cc/paper/2021/file/639d79cc857a6c76c2723b7e014fccb0-Paper.pdf}

\vspace{3mm}\noindent 
Abowd J, Ashmead R, Cumings-Menon R, Garfinkel S, Heineck M, Heiss C, Johns R, Kifer D, Leclerc P, Machanavajjhala A, Moran B, Sexton W, Spence M,  Zhuravlev P. 2022. 
The 2020 Census Disclosure Avoidance System TopDown Algorithm. 
\textit{Harv. Data Sci. Rev.} Spec. Iss. 2. 
\url{https://doi.org/10.1162/99608f92.529e3cb9 } 

\vspace{3mm}\noindent 
Abowd J, Adams T, Ashmead R, Darais D, Dey S, Garfinkel S, Goldschlag N, Kifer D, Leclerc P, Lew E, Porter E, Rodriguez R, Tadros R, Vilhuber L. 
Manuscript under review.
\textit{The U.S. Census Bureau’s Ex Post Confidentiality Analysis of the 2010 Census Data Publications}. 

\vspace{3mm}\noindent 
Abowd J, Schmutte I. 2015. 
Economic analysis and statistical disclosure limitation. 
In \textit{Brookings Papers on Economic Activity: Spring 2015}, ed. DH Romer, J Wolfers, pp. 221–-267. Washington, DC: Brookings Inst. 
\url{https://www.brookings.edu/wp-content/uploads/2015/03/AbowdText.pdf} cited on May 29, 2022

\vspace{3mm}\noindent 
Abowd J, Schmutte I. 2019. 
An economic analysis of privacy protection and statistical accuracy as social choices. 
\textit{Am. Econ. Rev.} 109(1):171-–202. 
\url{https://doi.org/10.1257/aer.20170627}

\vspace{3mm}\noindent 
Apple Differ. Priv. Team. 2017.  
Learning with privacy at scale. 
\textit{Apple Machine Learning Research,} Dec. 
\url{https://machinelearning.apple.com/research/learning-with-privacy-at-scale} cited on April 9, 2022


\vspace{3mm}\noindent 
Ashmead R. 2019. 
\textit{Estimating the variance of complex differentially private algorithms.} 
Presentation to the 2019 Joint Statistical Meetings, Denver, CO, July 27-Aug. 1. 
\url{https://www.census.gov/content/dam/Census/newsroom/press-kits/2019/jsm/presentation-estimating-the-variance-of-complex-differentially-private-algorithms.pdf} cited on April 9, 2022

\vspace{3mm}\noindent 
Balle B, Barthe G, Gaboardi M, Hsu J, Sato T. 
2020. 
Hypothesis testing interpretations and R\'enyi differential privacy. 
\textit{PMLR} 108:2496-2506. 
\url{http://proceedings.mlr.press/v108/balle20a.html} cited on April 8, 2022

\vspace{3mm}\noindent 
Benedetto G, Stanley JC, Totty E. 2018. 
\textit{The creation and use of the SIPP synthetic beta v7.0.} Work. Pap., US Census Bur., US Dep. Commer., Washington, DC
\url{https://www.census.gov/content/dam/Census/programs-surveys/sipp/methodology/SSBdescribe_nontechnicalv7.pdf} cited on April 11, 2022

\vspace{3mm}\noindent 
Bittau A, Erlingsson Ú, Maniatis P, Mironov I, Raghunathan A, et al. 2017. 
Prochlo: Strong privacy for analytics in the crowd. 
in \textit{SOSP '17: Proceedings of the 26th Symposium on Operating Systems Principles,} pp. 441--59. New York: ACM 
\url{https://doi.org/10.1145/3132747.3132769}

\vspace{3mm}\noindent 
Brenner H, Nissim K. 2014. 
Impossibility of differentially private universally optimal mechanisms. 
\textit{SIAM J Comput.} 43:1513--40.
\url{https://doi.org/10.1137/110846671}

\vspace{3mm}\noindent 
Bun M, Steinke T. 2016. 
Concentrated differential privacy: Simplifications, extensions, and lower bounds. 
In \textit{Theory of Cryptography (TCC 2016)} pp. 635–58. New York: Springer. 
\url{https://doi.org/10.1007/978-3-662-53641-4_24}

\vspace{3mm}\noindent 
Canonne C, Kamath G, Steinke T. 2020. 
The discrete Gaussian for differential privacy. 
In \textit{34th Conference on Neural Information Processing Systems (NeurIPS 2020)}, pp. 15676--88. N.p.:NeurIPS. 
\url{https://proceedings.neurips.cc/paper/2020/file/b53b3a3d6ab90ce0268229151c9bde11-Paper.pdf} cited on April 8, 2022

\vspace{3mm}\noindent 
Canonne C, Kamath G, Steinke T. 2021. 
The discrete Gaussian for differential privacy (extended version). 
\url{https://arxiv.org/abs/2004.00010}

\vspace{3mm}\noindent 
Childs J, Eggleston C, Fobia A. 2020. 
\textit{Measuring privacy and accuracy concerns for 2020 Census data dissemination.} Presented at BigSurv 20 Conference, online, Nov. 6--Dec. 4. 
\url{https://www.bigsurv.org/bigsurv20/uploads/25/82/Childs_BigSurv20_Paper_10.15.2020.pdf} cited on April 8, 2022.

\vspace{3mm}\noindent 
Comm. Natl. Stat. 2020.
\textit{2020 Census Data Products: Data Needs and Privacy Considerations: Proceedings of a Workshop.} Washington, DC: Natl. Acad. Press. 
\url{https://sites.nationalacademies.org/DBASSE/CNSTAT/DBASSE_196518?} cited on March 31, 2022

\vspace{3mm}\noindent 
Congr. Res. Serv. 2017. 
\textit{Congressional redistricting law: background and recent court rulings.} CRS Rep., Congr. Res. Serv., Washington, DC. 
\url{https://crsreports.congress.gov/product/pdf/R/R44798} cited on April 11, 2022

\vspace{3mm}\noindent 
Cox L. 1976. 
\textit{Statistical disclosure in publication hierarchies.} 
Rep. 14, Res. Proj. Confid. Surv., Dep. Stat., Univ. Stockholm, Swed.
\url{https://hdl.handle.net/1813/111306}

\vspace{3mm}\noindent 
Cox L. 1995. 
Network models for complementary cell suppression. 
\textit{JASA} 90(432):1453--62. 
\url{https://doi.org/10.2307/2291538}

\vspace{3mm}\noindent 
Dalenius T. 1977. 
Towards a methodology for statistical disclosure control. 
\textit{Stat. Tidskrift} 15:429--44. 
\url{https://hdl.handle.net/1813/111303}

\vspace{3mm}\noindent 
De Montjoye Y, Hidalgo C, Verleysen M, Blondel V. 
2013. 
Unique in the crowd: The privacy bounds of human mobility. 
\textit{Sci. Rep.} 3:1376. 
\url{https://doi.org/10.1038/srep01376}

\vspace{3mm}\noindent 
De Montjoye Y, Radaelli L, Singh VK, Pentland A. 
2015. 
Unique in the shopping mall: on the reidentifiability of credit card metadata. 
\textit{Science} 347(6221):536-39. 
\url{https://doi.org/10.1126/science.1256297}

\vspace{3mm}\noindent 
Devine J, Spence M. 2022. 
\textit{Update on the Demographic Profile and Demographic and Housing Characteristics file (DHC).} 
Presentation to the Census Scientific Advisory Committee, online, March 17. 
\url{https://www2.census.gov/about/partners/cac/sac/meetings/2022-03/presentation-demographic-profile-and-demographic-and-housing-characteristics.pdf} cited on April 9, 2022

\vspace{3mm}\noindent 
Ding B, Kulkarni J, Yekhanin S. 2017. 
Collecting telemetry data privately. 
In \textit{31st Conference on Neural Information Processing Systems (NIPS 2017)}. N.p.: NIPS. 
\url{https://papers.nips.cc/paper/2017/file/253614bbac999b38b5b60cae531c4969-Paper.pdf} cited on April 9, 2022

\vspace{3mm}\noindent 
Dinur I, Nissim K. 2003. 
Revealing information while preserving privacy. 
In \textit{PODS '03: Proceedings of the Twenty-Second ACM SIGMOD-SIGACT-SIGART Symposium on Principles of Database Systems,} pp. 202--10. New York: ACM. 
\url{https://doi.org/10.1145/773153.773173}

\vspace{3mm}\noindent 
Dong J, Roth A, Su WJ. 2020. 
Gaussian differential privacy.
\textit{J. R. Stat. Soc. Ser.  B} 86(1):3--37.  
\url{https://doi.org/10.1111/rssb.12454}, 
See also: \textit{Computing Research Repository (CoRR)}. Arxiv. \url{https://arxiv.org/abs/1905.02383} and \url{https://arxiv.org/abs/2104.01987}

\vspace{3mm}\noindent 
DSEP (Data Steward. Exec. Policy Comm.). 2010. 
\textit{January 14, 2010 final DSEP meeting record.} Rep., US Census Bur., US Dep. Commer., Washington, DC. 
\url{https://www.census.gov/about/policies/foia/foia_library/frequently_requested_records.html} cited on March 31, 2022

\vspace{3mm}\noindent 
DSEP (Data Steward. Exec. Policy Comm.). 2017. 
\textit{May 10, 2017 final DSEP meeting record.} Rep., US Census Bur., US Dep. Commer., Washington, DC. 
\url{https://www.census.gov/about/policies/foia/foia_library/frequently_requested_records.html} cited on March 31, 2022

\vspace{3mm}\noindent 
Duncan G, Elliot M, Salazar-Gonz\'alez JJ. 2011. 
\textit{Statistical Confidentiality: Principles and Practice}. New York: Springer. 
\url{https://doi.org/10.1007/978-1-4419-7802-8}

\vspace{3mm}\noindent 
Dwork C. 2006. 
Differential privacy. 
In: \textit{International Colloquium on Automata, Languages and Programming (ICALP 2006)} pp. 1--12. New York: Springer.
\url{https://doi.org/10.1007/11787006_1}

\vspace{3mm}\noindent 
Dwork C, McSherry F, Nissim K, Smith A, 2006a. 
Calibrating noise to sensitivity in private data analysis. 
In: \textit{Proceedings of the Theory of Cryptography Conference (TCC 2006)}, ed. S Halevi, T Rabin, pp. 265–-84. New York: Springer. 
\url{https://doi.org/10.1007/11681878_14}

\vspace{3mm}\noindent 
Dwork C, Kenthapadi K, McSherry F, Mironov I, Naor M. 2006b. 
Our data, ourselves: privacy via distributed noise generation. 
In: \textit{Annual International Conference on the Theory and Applications of Cryptographic Techniques (EUROCRYPT 2006)}, ed. S Vaudenay, pp. 486-503. New York: Springer. 
\url{https://doi.org/10.1007/11761679_29} 

\vspace{3mm}\noindent 
Dwork C, Naor M. 2010. 
On the difficulties of disclosure prevention in statistical databases or the case for differential privacy. 
\textit{J. Priv. Conf.} 2(1)93--107. 
\url{https://doi.org/10.29012/jpc.v2i1.585}

\vspace{3mm}\noindent 
Dwork C, Pottenger R. 2013. 
Toward practicing privacy. 
\textit{J. Am. Med. Inform. Assoc.} 20(1):102-–8.  
\url{https://doi.org/10.1136/amiajnl-2012-001047}

\vspace{3mm}\noindent 
Dwork C, Rothblum GN. 2016. 
Concentrated differential privacy. 
\url{https://arxiv.org/abs/1603.01887} [cs.DS]

\vspace{3mm}\noindent 
Dwork C, Smith A, Steinke T, Ullman J, Vadhan S. 2015. 
Robust traceability from trace amounts. Presented at the 56th Annual Symposium on Foundations of Computer Science (FOCS 2015). Oct. 18--20, Berkeley, CA. 
\url{https://doi.org/10.1109/FOCS.2015.46} 

\vspace{3mm}\noindent 
Dwork C, Yekhanin S. 2008. 
New efficient attacks on statistical disclosure control mechanisms. 
In \textit{CRYPTO 2008: Proceedings of the 28th Annual conference on Cryptology}, ed. D Wagner, pp. 469–-80. New York: Springer. 
\url{https://doi.org/10.1007/978-3-540-85174-5_26}

\vspace{3mm}\noindent 
Elamir E, Skinner C. 2004.  
\textit{Record-level measures of disclosure risk for survey microdata.} S$^{3}$RI Methodol. Work. Pap. M04/02, Southampton Stat. Sci. Res. Inst., Southampton, UK. 
\url{https://eprints.soton.ac.uk/8175/1/8175-01.pdf} cited on March 31, 2022


\vspace{3mm}\noindent 
Fellegi IP. 1972. 
On the question of statistical confidentiality. 
\textit{JASA} 67:7--18. 
\url{https://www.jstor.org/stable/2284695}

\vspace{3mm}\noindent 
Garfinkel S. 2015. 
\textit{De-identification of personal information.} Tech. Rep. NISTIR 8053, Natl. Inst. Standards Technol., Gaithersburg, MD. 
\url{https://csrc.nist.gov/publications/detail/nistir/8053/final#:~:text=Abstract,archiving\%2C\%20distributing\%20or\%20publishing\%20information.} cited on April 28, 2022

\vspace{3mm}\noindent 
Garfinkel S, Abowd J, and Martindale C. 2019. 
Understanding database reconstruction attacks on public data. 
\textit{Commun. ACM} 62(3):46--53. 
\url{https://doi.org/10.1145/3287287}

\vspace{3mm}\noindent 
Hammond EC, Horn D. 1954. 
The relationship between human smoking habits and death rates: a follow-up study of 187,866 men. 
\textit{JAMA} 155(15):1316--28. 
\url{https://doi.org/10.1001/jama.1954.03690330020006}

\vspace{3mm}\noindent 
Hawes M. 2020. 
Implementing differential privacy: seven lessons from the 2020 United States Census. 
\textit{Harv. Data Sci. Rev.} 2(2). 
\url{https://doi.org/10.1162/99608f92.353c6f99}


\vspace{3mm}\noindent 
Hawes M. 2022. 
\textit{Reconstruction and re-identification of the Demographic and Housing Characteristics File (DHC).} Presentation to the Census Scientific Advisory Committee, online, March 17. 
\url{https://www2.census.gov/about/partners/cac/sac/meetings/2022-03/presentation-reconstruction-and-reidentification-of-the-dhc.pdf} cited on April 8, 2022

\vspace{3mm}\noindent 
Hawes M, Rodriguez R. 2021. 
\textit{Determining the privacy-loss budget: research into alternatives to differential privacy.} Presentation to the Census Scientific Advisory Committee, online, May 25. 
\url{https://www2.census.gov/about/partners/cac/sac/meetings/2021-05/presentation-research-on-alternatives-to-differential-privacy.pdf} cited on April 9, 2022

\vspace{3mm}\noindent 
Homer N, Szelinger S, Redman M, Duggan D, Tembe W, Muehling J, Pearson J, Stephan D, Nelson S, Craig D. 2008. 
Resolving individuals contributing trace amounts of DNA to highly complex mixtures using high-density SNP genotyping microarrays. 
\textit{PLOS Genet.} 4:1--9. 
\url{https://doi.org/10.1371/journal.pgen.1000167}

\vspace{3mm}\noindent 
JASON. 2020. 
\textit{Formal privacy methods for the 2020 Census.} 
Distrib. Statement A, JSR-19-2F, JASON, Mitre Corp., Mclean, VA. 
\url{https://www.census.gov/programs-surveys/decennial-census/decade/2020/planning-management/plan/planning-docs/privacy-methods-2020-census.html} cited on April 9, 2022

\vspace{3mm}\noindent 
Johnson N, Near JP, Song D. 2018. 
Towards practical differential privacy for SQL queries. 
\textit{VLDB} 11(5):526--53. 
\url{http://vldb.org/pvldb/vol11/p526-johnson.pdf} cited on April 9, 2022

\vspace{3mm}\noindent 
Kasiviswanathan S, Smith A. 2014. 
On the `semantics' of differential privacy: a Bayesian formulation. 
\textit{J. Priv. Conf.} 6(1):1--16. 
\url{https://doi.org/10.29012/jpc.v6i1.634}


\vspace{3mm}\noindent 
Kim HJ, Drechsler J, Thompson KJ. 2020. 
Synthetic microdata for establishment surveys under informative sampling. 
\textit{J. R. Stat. Soc. Ser. A} 184:1:255--81. 
\url{https://doi.org/10.1111/rssa.12622}

\vspace{3mm}\noindent 
Kinney SK, Reiter JP, Reznek AP, Miranda J, Jarmin RS, Abowd JM. 2011. 
\textit{Towards unrestricted public use business microdata: the synthetic longitudinal business database.} Work. Pap. CES-11-04, US Census Bur., US Dep. Commer., Washington, DC.
\url{https://www.census.gov/library/working-papers/2011/adrm/ces-wp-11-04.html} cited on April 11, 2022

\vspace{3mm}\noindent 
Little RJA. 1993. 
Statistical analysis of masked data. 
\textit{J. Off. Stat.} 9(2):407-407. 
\url{https://www.scb.se/contentassets/ca21efb41fee47d293bbee5bf7be7fb3/statistical-analysis-of-masked-data.pdf} cited on April 11, 2022

\vspace{3mm}\noindent 
Little RJA, Rubin D. 2002
\textit{Statistical analysis with missing data} New York: Wiley. $2^{nd}$ ed.
\url{https://doi.org/10.1002/9781119013563}

\vspace{3mm}\noindent 
Ma CYT, Yau DKY, Yip NK, Rao NSV. 2013. 
Privacy vulnerability of published anonymous mobility traces. 
\textit{IEEE/ACM Trans. Netw.} 21(3):720--33. 
\url{https://doi.org/10.1109/TNET.2012.2208983}

\vspace{3mm}\noindent 
Machanavajjhala A, Kifer D, Abowd J, Gehrke J, Vilhuber L, 2008. 
Privacy: theory meets practice on the map. 
In \textit{Proceedings of the 2008 IEEE 24th International Conference on Data Engineering,} pp. 277-–86. New York: IEEE. 
\url{https://doi.org/10.1109/ICDE.2008.4497436}

\vspace{3mm}\noindent 
Massell D. 2001
Chapter VIII: the theory and practice of using data to build capacity: state and local strategies and their effects.
\textit{Teach. Coll. Rec.} 103(8)148-69
\url{https://doi.org/10.1177/016146810110300808}


\vspace{3mm}\noindent 
Meng X-L. 2020. 
2020: a very busy year for data science (and for HDSR). 
\textit{Harv. Data Sci. Rev.} 2(1). 
\url{https://doi.org/10.1162/99608f92.2ce040dc}

\vspace{3mm}\noindent 
McKenna L. 2018. 
\textit{Disclosure avoidance techniques used for the 1970 through 2010 Decennial Censuses of Population and Housing.} Work. Pap., US Census Bur., US Dep. Commer., Washington, DC. 
\url{https://www2.census.gov/ces/wp/2018/CES-WP-18-47.pdf} cited on April 10, 2022

\vspace{3mm}\noindent 
McKenna L. 2019. 
\textit{Disclosure avoidance techniques used for the 1960 through 2010 Census.} Work. Pap., US Census Bur., US Dep. Commer. Washington, DC.
\url{https://www.census.gov/library/working-papers/2019/adrm/six-decennial-censuses-da.html} cited on March 31, 2022

\vspace{3mm}\noindent 
Mironov I. 2017. 
R\'enyi differential privacy. In \textit{Proceedings of the 2017 IEEE 30th Computer Security Foundations Symposium CSF}, pp. 263--75. New York: IEEE. 
\url{https://doi.org/10.1109/CSF.2017.11}

\vspace{3mm}\noindent 
Narayanan A, Shmatikov V. 2008. 
Robust de-anonymization of large sparse datasets. In \textit{2008 IEEE Symposium on Security and Privacy (sp 2008)}, pp. 111--25. New York: IEEE. 
\url{https://doi.org/10.1109/SP.2008.33} cited on April 11, 2022

\vspace{3mm}\noindent 
NASEM (Natl. Acad. Sci. Eng. Med.) 2021. 
\textit{Principles and Practices for a Federal Statistical Agency.} Washington, DC: Natl. Acad. Press. 7th ed. 
\url{https://doi.org/10.17226/25885}

\vspace{3mm}\noindent 
NIH (Natl. Inst. Health). 2008. 
\textit{Modifications to GWAS data access.} Press Release, Aug. 28. 
\url{https://osp.od.nih.gov/wp-content/uploads/Data_Sharing_Policy_Modifications.pdf} cited on March 21, 2022

\vspace{3mm}\noindent 
OFSPS (Off. Fed. Stat. Policy and Stand.). 1978. 
\textit{Report on statistical disclosure and disclosure avoidance techniques.} Stat. Policy Work. Pap. 2, Off. Fed. Stat. Policy Stand., Off. Manag. Budg., Washington, DC. 
\url{https://nces.ed.gov/FCSM/pdf/spwp2.pdf} cited on March 31, 2022

\vspace{3mm}\noindent 
OMB (Off. Manag. Budg.). 1978.
Statistical policy directive no. 15: standards for the classification of federal
data on race and ethnicity. \textit{Fed. Regist.} 43(87):19269–70
\url{https://archives.federalregister.gov/issue_slice/1978/5/4/19259-19273.pdf#page=11} cited on November 29, 2022

\vspace{3mm}\noindent 
OMB (Off. Manag. Budg.). 1997. 
Statistical policy directive no. 15: revisions to the standards for the classification of federal data on race and ethnicity. \textit{Fed. Regist.} 62:58782.
\url{https://obamawhitehouse.archives.gov/omb/fedreg_notice_15/} 
\url{https://www.federalregister.gov/documents/1997/10/30/97-28653/revisions-to-the-standards-for-the-classification-of-federal-data-on-race-and-ethnicity?msclkid=9c958592b9d311ec8a4c9907e4294a42} cited on April 11, 2022

\vspace{3mm}\noindent 
OMB (Off. Manag. Budg.). 2014. 
Statistical policy directive no. 1: fundamental responsibilities of federal statistical agencies and recognized statistical units. 
\textit{Fed. Regist.} 79:71609--16. 
\url{https://www.federalregister.gov/documents/2014/12/02/2014-28326/statistical-policy-directive-no-1-fundamental-responsibilities-of-federal-statistical-agencies-and}

\vspace{3mm}\noindent 
Reamer A. 2020. 
Counting for dollars 2020: the role of the decennial Census in the geographic distribution of federal funds. 
Res. Rep., GW Inst. Public Policy, George Washington Univ., Washington, DC.
\url{https://gwipp.gwu.edu/counting-dollars-2020-role-decennial-census-geographic-distribution-federal-funds} cited on March 31, 2022

\vspace{3mm}\noindent 
Rodriguez R. 2021. 
\textit{Disclosure avoidance and the American Community Survey.} 
Presentation to the 2021 ACS Data Users Conference, online, May 20. 
\url{https://acsdatacommunity.prb.org/m/2021-acs-conference-files/147/download} cited on April 11, 2022

\vspace{3mm}\noindent 
Rubin DB. 1993. 
Statistical disclosure limitation. 
\textit{J. Off. Stat.} 9(2):461--68. 
\url{https://www.scb.se/contentassets/ca21efb41fee47d293bbee5bf7be7fb3/discussion-statistical-disclosure-limitation2.pdf} cited on April 11, 2022

\vspace{3mm}\noindent 
Steel P. 2013.
The Census Bureau's new cell suppression system.
In \textit{Proceedings of the 2013 Federal Committee on Statistical Methodology (FCSM) Research Conference.} Washington, DC: OMB.
\url{https://nces.ed.gov/FCSM/pdf/E3_Steel_2013FCSM-508.pdf} cited on November 29, 2022

\vspace{3mm}\noindent 
Sullivan T. 2020. 
Coming to our census: how social statistics underpin our democracy (and republic). 
\textit{Harv. Data Sci. Rev.} 2(1). 
\url{https://doi.org/10.1162/99608f92.c871f9e0}

\vspace{3mm}\noindent 
Thompson KJ, Kim H, Bassel N, Bembridge K, Coleman C, Freiman M, Garcia M, Kaputa S, Riesz S, Singer P, Valentine E, White TK, Whitehead D. 2020. 
\textit{Final report: Economic Census Synthetic Data Project research team.} USCB ADEP Work. Pap. Ser. ADEP-WP-2020-05, US Census Bur., US Dep. Commer., Washington, DC. 
\url{https://www.census.gov/content/dam/Census/library/working-papers/2020/econ/ADEP_WP_2020_05Jennyspaper.pdf?msclkid=30da8211b99311eca261e04115948bc9} cited on April 11, 2022

\vspace{3mm}\noindent 
\textit{Thornburg v. Gingles,} 478 U.S. 30 (1986).
\url{https://supreme.justia.com/cases/federal/us/478/30/} cited on May 28, 2022

\vspace{3mm}\noindent 
US House. 1790. \textit{Remarks by James Madison on the Bill for the 1790 Census.} 1st Congr., 2nd Sess., Feb. 2. 
\url{https://www.census.gov/programs-surveys/sis/resources/historical-documents/james-madison.html} cited on March 31, 2022

\vspace{3mm}\noindent 
USCB (US Census Bur.). 1994. 
\textit{Geographic Areas Reference Manual}. Washington, DC: US Dep. Commer. 
\url{https://www2.census.gov/geo/pdfs/reference/GARM/} cited on April 9, 2022

\vspace{3mm}\noindent 
USCB (US Census Bur.). 2011. 
\textit{2010 Census Redistricting Data (Public Law 94-171) Summary File: Technical Documentation.} Washington, DC: US Dep. Commer. 
\url{http://www2.census.gov/programs-surveys/decennial/rdo/about/2010-census-programs/2010Census_pl94-171_techdoc.pdf} cited on March 31, 2022

\vspace{3mm}\noindent 
USCB (US Census Bur.). 2015.
\textit{United States Public Use Microdata Sample (PUMS): Technical Documentation.} Washington, DC: US Dep. Commer.
\url{https://www2.census.gov/programs-surveys/decennial/2010/technical-documentation/complete-tech-docs/us-pums/pumsus.pdf} cited on March 31, 2022

\vspace{3mm}\noindent 
USCB (US Census Bur.). 2019. 
\textit{The 2020 Census and Confidentiality.} Fact Sheet, US Census Bur., US Dep. Commer., Washington, DC.
\url{https://www.census.gov/content/dam/Census/library/factsheets/2019/comm/2020-confidentiality-factsheet.pdf} cited on April 8, 2022


\vspace{3mm}\noindent 
USCB (US Census Bur.). 2021a. 
\textit{2020 Census state redistricting data (Public Law 94-171) summary file: technical documentation.} Summary File, US Census Bur., US Dep. Commer., Washington, DC.
\url{https://www2.census.gov/programs-surveys/decennial/2020/technical-documentation/complete-tech-docs/summary-file/2020Census_PL94_171Redistricting_StatesTechDoc_English.pdf} cited on April 9, 2022

\vspace{3mm}\noindent 
USCB (US Census Bur.). 2021b. 
\textit{Developing the DAS: demonstration data and progress metrics;2020 Census redistricting data (P.L. 94-171), production settings: 2021-06-08: detailed summary metrics}. Program Inf.,  US Census Bur., US Dep. Commer., Washington, DC.
\url{https://www.census.gov/programs-surveys/decennial-census/decade/2020/planning-management/process/disclosure-avoidance/2020-das-development.html} cited on March 31, 2022

\vspace{3mm}\noindent 
USCB (US Census Bur.). 2022a. 
\textit{Understanding disclosure avoidance-related variability in the 2020 Census redistricting data.} Fact Sheet, US Census Bur., US Dep. Commer., Washington, DC.
\url{https://www.census.gov/library/fact-sheets/2022/variability.html} 
and \url{https://www.census.gov/library/working-papers/2021/adrm/CED-WP-2021-007.html} cited on April 29, 2022

\vspace{3mm}\noindent 
USCB (US Census Bur.). 2022b. 
\textit{LEHD Origin-Destination Employment Statistics (2002-2019)} Data Set, US Census Bur., US Dep. Commer., Washington, DC.
\url{https://onthemap.ces.census.gov} [version] LODES 7.5 cited on April 9, 2022

\vspace{3mm}\noindent 
Wasserman L, Zhou S. 2010. 
A statistical framework for differential privacy. 
\textit{JASA} 105:375--89. 
\url{https://doi.org/10.1198/jasa.2009.tm08651}

\vspace{3mm}\noindent 
Willenborg L, de Waal T. 2001. 
\textit{Elements of Statistical Disclosure Control} New York: Springer. 
\url{https://doi.org/10.1007/978-1-4613-0121-9}

\vspace{3mm}\noindent 
Wright T, Irimata K. 2021. 
\textit{Empirical study of two aspects of the Topdown Algorithm output for redistricting: reliability \& variability (August 5, 2021 Update)} Rep., USCB CSRM Studies Ser. Stat. \#2021-02, US Census Bur., US Dep. Commer., Washington, DC. 
\url{https://www.census.gov/content/dam/Census/library/working-papers/2021/adrm/SSS2021-02.pdf} cited on March 31, 2022

\vspace{3mm}\noindent 
Yu F, Fienberg SE, Slavkovi\'c A, Uhler C. 2014. 
Scalable privacy-preserving data sharing methodology for genome-wide association studies. 
\textit{J. Biomed. Inform.} 50:133--41. 
\url{https://doi.org/10.1016/j.jbi.2014.01.008}


\end{document}